# Class invariants: concepts, problems and solutions

**Bertrand Meyer**

*Draft 4, 22 May 2021*

*Groucho: That's in every contract. That's what they call a sanity clause.*
*Chico: Haha! You can't fool me. There ain't no sanity clause.*

*From* [19]

## Abstract

Class invariants are both a core concept of object-oriented programming and the source of the two key open OO verification problems: *furtive access* (particularly from callbacks) and *reference leak* (from aliasing). Existing approaches force on programmers an unacceptable annotation burden. This article explainfps the concept of invariant and solves both problems modularly through the *O-rule*, defining fundamental OO semantics, and the *inhibition rule*, using information hiding to remove harmful reference leaks. It also introduces the concept of *tribe* as a basis for other possible approaches.

*For all readers*: this article is long because it includes a tutorial, covers many examples and dispels misconceptions. To understand the key ideas and results, however, the first page-and-a-half (section 1) suffices.

*For non-experts in verification*: all concepts are explained; anyone with a basic understanding of object-oriented programming can understand the discussion.

*For experts*: the main limitation of this work is that it is a paper proposal (no soundness proof, no implementation). It addresses, however, the known problems with class invariants, solving such examples as linked lists and the Observer pattern, through a simple theory and without any of the following: ownership; separation logic; universe types [34]; object wrapping and unwrapping [15]; semantic collaboration, observer specifications [39, 40]; history invariants [17]; "inc" and "coop" constructs [31]; friendship construct [3]; non-modular reasoning [25]. More generally, it involves no new language construct and no new programmer annotations.

# 1 Overview and main results

In object-oriented programming, every class is characterized by a class invariant: a sanity clause expressing that its instances are compatible with the abstract purpose of the class.

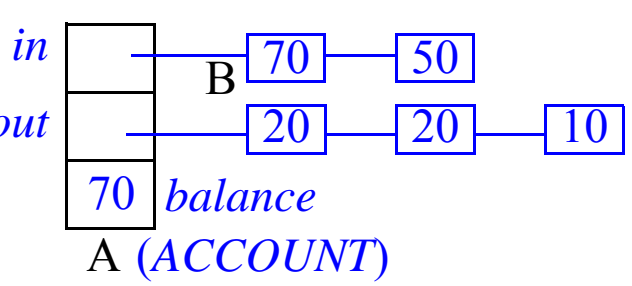

Assume that a bank account class specifies lists *in* and *out* of deposits and withdrawals and a value *balance*. Not every combination of these represent a meaningful account: they must satisfy $balance = in.total - out.total$ where *total* gives the accumulated value of a list of operations. This property is a class invariant.

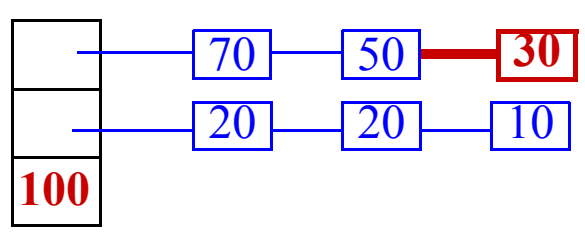

Per OO rules, objects change as a result of qualified calls such as $a.deposit\ (30)$. Such a call may assume on entry that the object satisfies the invariant and must ensure on exit that it satisfies it again. The routine *deposit* must both update the value of *balance* (here increasing it by 30) and add to the *in* list an object (with value 30) representing the withdrawal operation.

Whether or not programmers use them or have even heard the term, class invariants are one of the fundamental concepts of OO programming. They are essential to proving the correctness of OO programs, but raise two problems which have not received satisfactory solutions so far. The contribution of this article is a proposed solution to both of them. The rest of this section sketches the problems and the solutions. The following sections provide background and details.

The first problem, furtive access, arises when the routine of a qualified call, here *deposit*, performs a *callback* into the original object and finds it in a temporary state that does not satisfy the invariant. The solution is the following "O-rule" for qualified calls (explanations follow):

$$/O/ \qquad \frac{\{INV_r \wedge Pre_r\ (f)\}\ body_r\ \{INV \wedge Post_r\ (f)\}}{\{INV_r \wedge x.Pre_r\ (a)\}\ \textbf{call}\ x.r\ (a)\ \{x.INV \wedge x.Post_r\ (a)\}}$$

Surprisingly, no generally accepted inference rule seemed until now to exist for the fundamental construct of OO programming: routine call (message passing) **call** $x.r$[1]. The O-rule fills this gap.

Notations: a routine $r$ with formal arguments $f$ has implementation $body_r$, precondition $Pre_r$ and postcondition $Post_r$. The instruction **call** $x.r\ (a)$ calls $r$ on a target $x$ with actual arguments $a$. $INV$ is the class invariant applied to the current object, $x.INV$ the invariant applicable to $x$. Finally, $INV_r$ is the part of $INV$ containing only clauses that involve features exported no more than $r$.

A rule of this kind is a permission to infer the conclusion (below the line) if you have established the hypothesis (above). *Proving a class correct* means proving the hypothesis for every routine of the class: the body, started with the precondition and partial invariant[2], yields the postcondition and the full invariant. A class is a reusable software component (once proved, a "trusted" component [24]) and can be used in qualified calls **call** $x.r\ (a)$ for $x$ of the corresponding type. The conclusion part of the rule tells us how to reason about such a call:

- Obligation: we *must* establish that the context before the call satisfies $INV_r \wedge x.Pre_r\ (a)$.
- Benefit: we *may* then deduce that the context after the call will satisfy $x.INV \wedge x.Post_r\ (a)$.

1. Modern programming languages omit the keyword **call**, added in a few places in this article for clarity.
2. For secret (non-exported) $r$, $INV_r$ is empty, so the invariant plays no role on the left.

The invariant plays two complementary roles:

- On the left, $INV_r$, expresses the sanity of the current object, which "clams up" to be ready for callbacks. The *export consistency rule* ensures that callbacks need no more than $INV_r$.
- On the right, $x.INV$ expresses the sanity of the target. Here we need the full invariant because we cannot predict which call will hit $x$ next, but it must regardless find the object in a sane state.

The big bonus of the O-rule is that on entry to a qualified call we never have to *establish* the invariant on the target ($x.INV_r$). The O-rule enables us to assume, even in the presence of callbacks, that all previous calls have preserved the invariant, therefore justifying this term.

If it can be established that $r$ will cause no callback, the rule simplifies to:

/O'/
$$\frac{\{INV_r \wedge Pre_r(f)\}\ body_r\ \{INV \wedge Post_r(f)\}}{\{x.Pre_r(a)\}\ \textbf{call}\ x.r(a)\ \{x.INV \wedge x.Post_r(a)\}}$$

making it easier to prove the correctness both of the class ($INV$ can be stronger than $INV_r$, so we can assume more) and of a call: no more need, on entry, to establish the invariant or any part of it.

The second problem, reference leak, arises when the invariant for an object A involves properties of an object B — said to **inhibit** A — but a third object C changes B, invalidating A's invariant. Here the *in* list (B) inhibits the account (A) since *ACCOUNT*'s invariant involves *in* (in *in.total*). An object C could hold a reference $l$ to the list and, through $l.extend$ (80), insert an object of value 80. Unlike *ACCOUNT*, C's class has no obligation to update A's *balance*: the call will break A's invariant, yielding an incorrect object structure even though each routine preserves its own class's invariant.

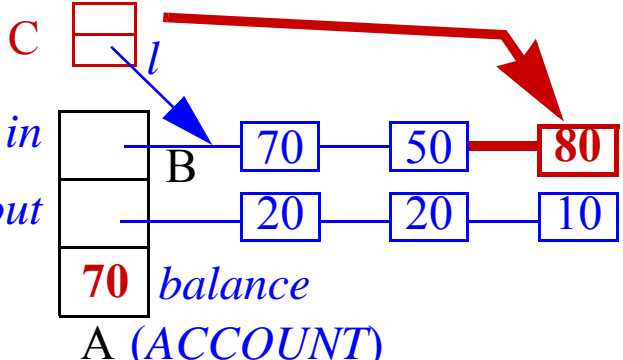

Previous approaches rely on "ownership", which requires heavy annotations and fails to address many practical cases. The new solution, the *inhibition rule*, is a simple addition to information hiding. It forces B's class to export operations affecting the inhibiting property (here *in*) to A's class only (*ACCOUNT*); then harmful leaks could only come from other instances of that class. To eliminate such leaks, the rule's second part simply prohibits A's class from exporting to *itself* any operation with an argument or result of B's type (the type of *in*). The inhibition rule even handles such cases of *mutual* inhibition as the Observer pattern, which ownership cannot handle.

We will now see the details. Section 2 introduces guidelines. Section 3 presents the concept of invariant and section 4 a set of examples causing difficulties. Section 5 highlights common misconceptions. Sections 6 to 8 discuss the furtive access problem and the two rules that address it: O-rule and Export Consistency. Section 9 covers reference leak and its resolution through the inhibition rule. Section 10 describes a more tentative approach: *tribes*. Section 11 discusses limitations.

## 2 Methodological guidelines

The following criteria guide this work.

The general goal is to build a software development environment applying the idea of "Verification As a Matter Of Course" [28]: programmers should be able to verify their programs as they develop them, treating verification not as a special step requiring extraordinary effort but as a normal part of the development process. These programmers should understand basic concepts but do not need to be experts in verification techniques.

Any verification effort requires that programmers add some annotations to their programs — you cannot verify programs without specifying what properties you expect them to satisfy — but the annotation effort should remain commensurate with the benefits. Concretely, we expect programmers to express the goal of every routine through a precondition and postcondition, and the characteristics of every class through the class invariant, but any further annotation demand is questionable[3]. The approaches proposed so far to address the problems of class invariants — whether using ownership (for example [3]), separation logic or semantic collaboration [39] — require programmers to specify many properties that do not pertain to the goal of the program but guarantee special conditions imposed by the verification technique. To turn verification into a matter of course, we must shield programmers from these expert-level concerns.

Another characteristic of the present work is that it follows a tradition of understanding "verification" as including both static and dynamic checks: proofs and tests. Recent literature often uses "verification" as a synonym for "proof", but the practice of Design by Contract also uses contract elements (preconditions, postconditions, invariants) as sanity conditions that can optionally monitored at execution time, as a tool for testing and debugging.

Today's work on verification attaches considerable importance to modularity (in the words of Leino and Müller in [16]: "*it should be possible to reason about smaller portions of a program at a time, say a class and its imported classes, without having access to all pieces of code in the program that use or extend the class*"). The techniques developed here fulfill this criterion.

## 3 Class invariants: history and tutorial

We now review the concept of class invariant, beginning with its history.

### 3.1 Origin

The concept comes from a 1972 paper by Hoare [12], which contains the first recorded occurrence of the term ("*invariant of the class*"). The paper considers a data type that has both an abstract specification and a particular implementation using variables $c_1, c_2, \dots c_n$[4]; in the example above the abstract concept is "bank account" and the variables are *in*, *out* and *balance*. Hoare writes:

> *For practical proofs we need a slightly stronger rule, which enables the programmer to give an invariant condition $I(c_1, c_2, \dots c_n)$, defining some relationship between the constituent concrete variables, and thus placing a constraint on the possible combinations of values which they may take. Each operation (except initialization) may assume that $I$ is true when it is first entered; and each operation must in return ensure that it is true on completion.*

In today's more general view of invariants, this description covers the special case of *representation* invariants. A few experimental 1970s languages included support for representation invariants; most notable is Alphard [41], which provides for both an abstract invariant (characterizing the abstract data type) and a representation invariant (characterizing a particular representation).

In object-oriented programming, inheritance subsumes this distinction. Abstraction and representation become relative concepts; each class inherits the invariants of its parents, adding its

3. *Loop* invariants [9] lie on the borderline. In the current state of verification technology, they are still hard to infer automatically.

4. Hoare's classic paper contains a small but interesting oversight: it talks about a "*representation function*" mapping abstract to concrete objects. This relation is not a function, since an abstract object may have many implementations. We do get a function if we consider its *inverse*: the *abstraction function*.

own. As a result, the refinement process may have any number of levels, rather than just two. This observation, and more generally the development of the concept of class invariant for object-oriented programming, appeared in 1985 in [20] and subsequent publications about Eiffel. [21] in 1988 and [23] in 1997 (hereafter called OOSC 1 and OOSC 2) explored the concepts further. A number of other verification-oriented formalisms have included support for class invariants, notably JML [6, 14] and Spec# [18].

The correctness-by-construction school of program development has also relied on a notion of invariant. In these approaches (Back [2], Morgan [32], Event-B [8]), system construction proceeds by refinement steps, starting from a abstracted high-level description of the system, to which every subsequent step adds more detail. The process reaches its final step when all the desired behavior elements are in place and can be directly implemented in a programming language. Constraining the description at every step is an invariant, and constraining the refinement to the next step is the obligation to preserve the previous invariant while adding invariant clauses governing the new details. These rules resemble the accumulation of invariant clauses in inheritance. Indeed, while refinement approaches do not use the rest of the object-oriented paradigm, refinement is essentially the same idea as inheritance, and the concept of invariant is the same.

## 3.2 The class invariant concept

The define the semantics of invariants we may use the following "*Fundamental Picture*", taken from OOSC (1 and 2).

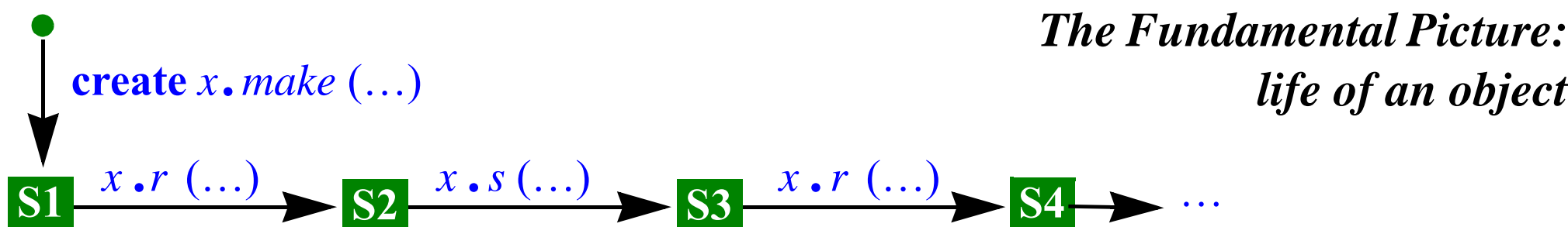

***The Fundamental Picture: life of an object***

It illustrates the life of an object. In truth, a boring life:

- At the beginning, someone — a routine executed on behalf of another object[5] — creates the object, using a creation procedure of the class, or "constructor", here *make* (the syntax in languages such as Java is *x* = **new** *T* (…) where *T* is the desired type).
- Ever after, all that happens to the object is that routines executed on behalf of other objects[6] execute qualified calls on it. Each of these calls applies to the object one of the routines declared in its class; in the figure: *r*, then *s*, then *r* again.

Such calls are said to be "qualified" because they apply a routine to a target object, here called *x*[7] and followed — in the syntax of most OO languages — by a dot. The meaning of this fundamental construct, *x.r (a, …)*, is: "apply the routine *r* to the object known as *x*, with the given actual arguments *a, …* if any".

The Fundamental Picture shows how the object goes from state to state (S1, S2 and son on) as a result of qualified calls.[8]

5. For the very first object in an execution, the "root" object, the trigger comes from some external mechanism.
6. Or the object itself, causing "qualified callbacks" as discussed next.
7. *x* is not an object but a name in the program denoting possible run-time objects. Different classes, and even different parts of a single class (because of aliasing), may use different names for the same object.
8. The picture applies to concurrent as well as sequential computation. The successive calls can come from different *processors* (threads of control, see [30]).

There is also a possibility of *unqualified* calls, which the object applies to itself. For example the routine *s* could be declared as

```
s (x: INTEGER)
    do
        u.do_unto_other (x + 1)        -- A qualified call on the object attached to u.
        …                              -- Possible other instructions.
        do_unto_me                     -- An unqualified call, on the current object.
    end
```

The Fundamental Picture only shows the states after creation and after qualified calls (each of which, except for the final one in an execution, is also a state *before* a qualified call). We may call them "sane states" of the object. In-between sane states, there may be many intermediate states, for example just before the call to *do_unto_me* above. Such an "intermediate state" is represented by a round dot in this refinement of the Fundamental Picture:

***An intermediate state***

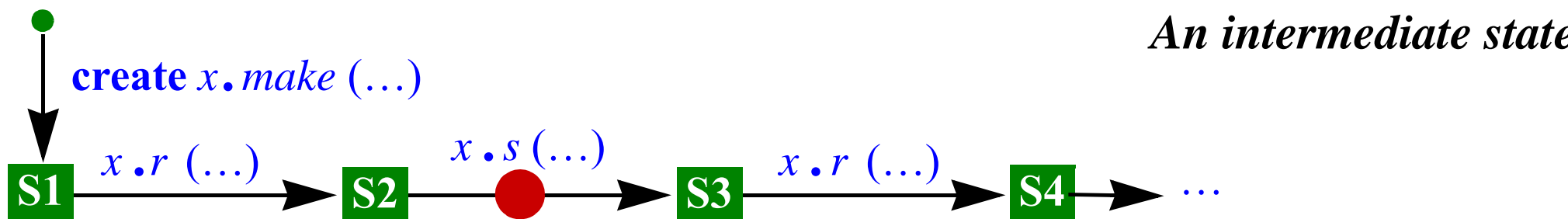

The class invariant characterizes sane states, those in which the object is available to other objects. Intermediate states do not need to satisfy it. In practice, most interesting computations do violate the invariant — if you try to do something useful, you often have to start by disturbing the established order — then restore it.

The following analogy from OOSC helps reason about class invariants and the Fundamental Picture. Think of a shared kitchen in an office environment, which has a sign enjoining users to "*make sure you leave this place as you want it to be when you come in*". This is what invariants are about. You *may* assume the invariant on entry (clean sink, ready-to-use coffee machine, …) and *must* ensure it on exit. "Ensure" here often means "restore": while using the kitchen, it is all right to mess it up, as in the red-dot state of the figure above, as long as you clean up your mess for the next user.

Taking the Fundamental Picture as a reference implies that we only need to verify that every routine of the class (such as *r* and *s* here) *preserves* the invariant, in the sense that if the invariant holds before an execution of the routine it will hold afterwards; also, that it will hold after the execution of every constructor (obviously, we do not assume that it holds before). Then in verifying client code that uses a qualified call *x.r* (…) we may deduce that the invariant applied to *x* (the notation, as the reader will remember, is *x.INV*); we should be entitled to this conclusion *without* having to establish that *x.INV* holds before the qualified call[9]. Herein lies the beauty and power of the notion of class invariant.

This article is devoted to studying how we can make this ideal scheme a reality.

## 3.3 Assumptions

Many practical cases, most in fact, do follow the Fundamental Picture. But it makes two implicit assumptions, which do not always hold:

- The picture assumes that qualified calls (the horizontal arrows) are computations on the target object. They may involve many tortuous steps, but all apply that object. To paraphrase a famous marketing slogan, what happens to the target stays within the target. In reality, these computations may themselves perform qualified calls; so they can modify, in addition to fields

9. [31] calls this property the "data induction theorem".

of that object, the contents of other objects. So far so good, but these computations on other object may *come back* to the original target, complicating the picture. This is a case of the *furtive access* problem.

- The picture also assumes that we have a single name, *x*, for the target object. But OO programming allows aliasing: different places in the program may know a given object under different names. They can mess up with each other in a way not captured by the picture: even if each operation preserves the invariant of its own class, it may break the invariant of an object that also depends on the shared object. This is a case of the *reference leak* problem.

Furtive access and reference leak are the two difficulties that we will have to analyze and address.

## 3.4 A simple example

The following pedagogical example of an invariant-equipped class is about as simple as one can get. The class describes points on a line, with an integer coordinate constrained to remain between 0 and a maximum value L, 4 in the following figure.

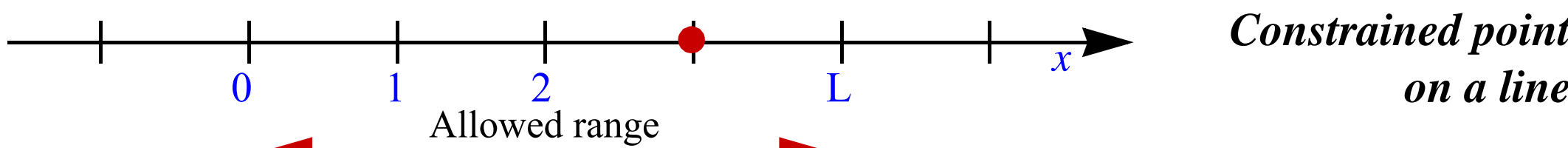

***Constrained point on a line***

We may write the class as:

```
class POINT feature
    x: INTEGER
    move_left do … See below … end
  invariant
      non_negative: 0 <= x
      bounded: x <= L
end
```

(The syntax makes it possible to document assertions with tags, such as non_negative, which help readability but have no semantic value.) The routine *move_left* has the following specification and implementation:

```
move_left
    do
        x := x – 1
        if x < 0 then x := L end        -- When falling off on the left, jump over to the right.
    ensure
        decreased_if_positive: (old x > 0) => (x = old x – 1)-- "=>" is logical implication.
        decreased_or_wrapped: (x = old x – 1) or (x = L)
    end
```

Note that with the specific implementation shown it would be possible, for the second postcondition clause decreased_or_wrapped, to use a stronger version:

```
        wrapped_around_if_zero: (old x = 0) => (x = L)
```

which more clearly expresses the intent: *x* either decreases by one if initially positive (decreased_if_positive), or gets set to L if initially zero (wrapped_around_if_zero). We intentionally leave the weaker version decreased_or_wrapped, because it illustrates the relevance of the invariant: combined with the invariant clause non_negative, the postcondition clause decreased_or_wrapped does imply wrapped_around_if_zero.

This specification — as provided by the postcondition and the invariant — makes it possible to prove that after the following creation and call, for *p* of type *POINT*, *p.x* will be equal to L:

```
create p            -- Uses default creation: sets p.x to 0.
p.move_left
```

The postcondition tells us that *p.x* is either –1 or L; then thanks to the invariant we can rule out the first possibility.

Although reduced to bare bones, this example illustrates the role of invariants in object-oriented programming as suggested by the Fundamental Picture.

Terminology: the term **object invariant** will denote the class invariant applied to a particular object. Here the object invariant of a particular point is the property that its *x* is between 0 and L.[10]

Class invariants are central to the OO method. They endow every class with a sanity clause guaranteeing that the class represents the expected abstraction. The concept reflects the notion of *axiom* in the underlying mathematical theory, abstract data types (OOSC explains the connection). In the teaching of OO programming, even at the elementary level, invariants deserve the same emphasis as other fundamental OO concepts: classes, information hiding, genericity, inheritance and contracts ([26] is an introductory programming textbook applying this idea). As Groucho Marx suggests, they should be in every contract and class. Regrettably, the reality is closer to Chico's view: in the classes most programmers write, there is no sanity clause.

## 4 Representative examples of invariant-related trouble

The invariant of class *POINT* showed the basic idea. The following examples illustrate the two problems, furtive access and reference leak, and will serve as a testbed for the solutions. Some are drawn from the literature on class invariant problems, some are new, but together they are representative of the combined set of examples used in previous discussions of these problems.

### 4.1 Unregistered observers

The first example introduces observer objects, each observing an instance of the class *POINT*, but in a primitive sense of the notion of observer: they are not automatically notified when the observed object changes. Such an object is like a simple "façade" to the corresponding point. The next example will have observers in the full sense of the Observer pattern. The example shows how easy it is to cause a *reference leak* as soon as aliasing occurs in the presence of inhibition (defined as the presence of invariant clauses of the form *some_object.some_property*).

Our observer objects will observe a point that is not at the right boundary L of its range[11]:

```
class UNREGISTERED_OBSERVER create set feature
    subject: POINT
    set (other: POINT)                  -- Initialize subject to other.
        require other.x < L do subject := other ensure subject = other end
    … Other features, all preserving the invariant …
invariant
    subject.x < L
end
```

10. Some articles use "object invariant" in a confusing way; see footnote 30.

11. L is an arbitrary positive constant. It can be replaced by its actual value, e.g. 100, in these examples.

The invariant states the desired property. Any exported feature of *UNREGISTERED_OBSERVER* must leave the *x* of *subject* to a value less than L. On initialization, *set* sets *subject* to an existing point, ensuring the invariant thanks to its precondition.

Although this class is simple and its semantics clear, difficulties will arise from the highlighted invariant clause *subject.x* < L which involves a query on another object, the *subject*. We say that this clause causes the subject to *inhibit* the observer.

Here now is another client of *POINT*, which also has a *subject* attribute but no particular condition on it, and a reference to an instance of *UNREGISTERED_OBSERVER*:

```
class MISCHIEF create make feature
    subject: POINT               -- "create subject", in make below, will initialize subject.x to 0.
    obs: UNREGISTERED_OBSERVER
    make do create subject ; create obs.set (subject) end
    mess_up do subject.move_left end
end
```

Let us create an instance of *MISCHIEF* through **create** *m.make*. The first instruction of the *make* creation procedure creates an instance of *POINT*; the second instruction creates an instance of *UNREGISTERED_OBSERVER*, with a reference to the same *POINT* object:

***Reference leak***

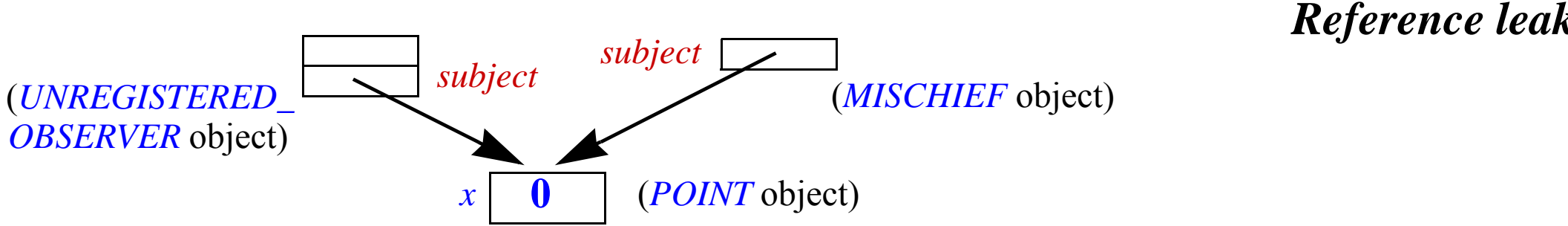

A reference leak has happened: the *MISCHIEF* object has leaked its *subject* reference to the *UNREGISTERED_OBSERVER* object. A call to *mess_up* will use the leaked reference to change the *x* field of the shared object, setting it to L and hence violating the invariant of *UNREGISTERED_OBSERVER*.

This case in archetypal example of the reference leak problem. The trouble is that in the absence of further rules no modular check of either class can detect this problem: *MISCHIEF* has no invariant; its argument in the creation call to *set* satisfies that routine's precondition (at that stage, *subject.x* is still zero); and all the routines of *UNREGISTERED_OBSERVER* preserve the invariant of that class. Each class, when studied by itself, appears correct; but together they are not, since *an object may become invalid without any operation being explicitly performed on it*.

## 4.2 Observer pattern

We move now to a true observer, in the sense of the Observer pattern [10]: an object that watches a point and gets updated automatically every time that point changes. That way, other objects can find out the location of a point by querying the observer. The class will look like this:

```
class OBSERVER create make feature
    x: INTEGER
feature {POINT}          -- See below about the meaning of "{POINT}" and (next) {NONE}".
    subject: POINT
    update do x := subject.x ensure x = subject.x end
feature {NONE}           -- Initialization
    make (p: POINT) do subject := p ; subject.set_observer (Current) end
            -- Set p as the observed point. See below about set_observer.
invariant
    faithful: x = subject.x                    -- For ease of reference an assertion clause can
    backlink: subject.observer = Current       -- have a tag, such as faithful and backlink here.
end
```

The observer scheme has been widely used as a justification for sophisticated language and verification constructs [3, 15, 31, 39, 40] and as a challenge in a verification competition [5][12].

Notation: this example uses the "selective export" mechanism, where a feature *r* introduced in a clause **feature** {*A*, *B*, *C*} is only available for qualified calls *x*.*r* (…) in the classes listed and their descendants. Here, *some_point*.*subject* and *some_point*.*update* are only permitted in *POINT* and descendants. Selective exports apply the principle of information hiding. The mechanism exists in many OO languages, in various syntactic flavors such as "friends" in C++ and "assembly" privileges in C# and Visual Basic.Net. A clause reading just **feature** is equivalent to **feature** {*ANY*}, where *ANY*[13] is the top class in the inheritance hierarchy: it introduces public features. A clause reading **feature** {*NONE*} where *NONE* (Object in Java) is the bottom class, introduces private (secret) features[14]. We could make *subject* and *update* public (like *x*), but good design methodology directs that we should not let clients other than *POINT* access them.

Back to invariants and the semantics of the class: any routine of *POINT* modifying *x* must now notify its associated observer. In our simplified example, there is only one such routine, *move_left*. Its body will now be as follows (first two lines unchanged, the only new element is the last line):

```
x := x – 1                   -- This is the x of POINT, observed by the x of OBSERVER
if x < 0 then x := L end
if observer ≠ Void then observer.update end        -- This is the added line.
```

Class *POINT* now needs knowledge of the observer, through a new attribute and routine:

```
feature {OBSERVER}
    observer: detachable[15] OBSERVER
    set_observer (o: OBSERVER) do observer := o end      -- Link point to observer o.
```

These features are selectively exported to *OBSERVER* since the association with the observer is none of the rest of the world's business (but we do need to let *OBSERVER* access *observer*, if only for its *backlink* invariant clause). Mirroring *OBSERVER*, class *POINT* now has a *backlink* invariant clause stating

---

12. Like practical uses of the Observer pattern, those articles assume a list of observers, rather than a single observer; since the list adds nothing to the issue, the example is given here reduced to its simplest form.

13. *ANY* is called Object in Java.

14. *make* is a creation procedure since it appears in the **create** clause, allowing creation-cum-initialization instructions **create** *obs*.*make* (*p1*). We could also permit ordinary qualified calls *obs*.*make* (*p1*), to reset the subject of an existing observer, but here they are disallowed since the feature is declared in a clause **feature** {*NONE*}.

(*observer* ≠ **Void**) => (*observer*.*subject* = **Current**)

Reference leak is possible here, just as with unregistered observers (because of "inhibiting" invariant clauses involving qualified calls such as *subject*.*x* and *observer*.*subject*). But this example also causes a new problem. Several articles using some form of it point out that the *observer*.*update* call highlighted above catches the observer object with its pants down (so to speak): the invariant clause *faithful* usually does not hold on entry, since it is precisely the purpose of the call to *update* to make sure that after a change *x* will again faithfully reflect the observed value *subject*.*x*.

This case illustrates the furtive access problem: accessing an object in a state that does not satisfy the corresponding class invariant. Furtive access also arises, as we will see, as a result of qualified callbacks.

To address the problem, the literature offers special language and verification constructs. But we may note that in this case the invariant violation *does not matter*. It occurs in an intermediate state of the object where it is harmless. (When you invite an office colleague to join a party already in progress, you are not promising that the kitchen is in order.) This article's solution to furtive access —the O-rule — will take advantage of this observation.

### 4.3 Cloning

The following example, like the observer case, illustrates furtive access.

It arose early in the design of the EiffelBase library [22]. Any class may offer its own version of the *copy*, *cloned* and *is_equal* routines. Default versions exist, pre-programmed, in the top class *ANY*: respectively, they copy an object onto another, create a new object as a duplicate of an existing one, and test two objects for field-by-field equality. The difference between *copy* and *cloned* is that *y*.*copy* (*x*) copies the contents of an existing object *x*[16] onto those of an existing object *y*, whereas *y* := *x*.*cloned* produces a new object *y* identical to *x*.

To adapt copy and equality semantics to a particular class (in a linked list class, for example, the routines should copy and compare entire lists, not just the list headers as the default versions do), you must redefine *copy* and *is_equal*, maintaining for *copy* (*x*) the postcondition *is_equal* (*x*).

You should not, however, have to redefine *cloned* in addition to *copy*. The semantics should be the same except for the creation of a new object; *cloned* should automatically follow *copy*. In the original version of the library the implementation of *cloned* directly applied this idea:

```
Result := Blank_object (Current_type)
Result.copy (Current)
```

with a system function *Blank_object* returning a zeroed-out freshly allocated object of the given type. The first time someone ran a program using *cloned* with run-time contract monitoring[17] on, the highlighted instruction violated an invariant. Indeed, a zeroed-out object will not satisfy any non-trivial invariant.That is precisely why we call *copy* to turn it into a sane object[18].

15. **detachable** indicates that *observer*, unlike *subject* in *OBSERVER*, can be void (null). This entire discussion assumes a void-safe language — one that guarantees statically that no null-pointer dereferencing will ever happen during the execution of any program. On how to achieve void safety see [27].

16. Small abuse of language for "the object denoted by *x*" etc.

17. Run-time contract monitoring evaluates invariants and other contract elements during execution, as an example of dynamic verification, the widely applied use of contracts so far until static verification technology becomes fully practical.

The last observations reveal how similar the cloning case is to our other example of furtive access, the observer case: a non-updated observer will usually not satisfy the invariant clause; that is precisely why we call *copy* to turn it into a sane observer.

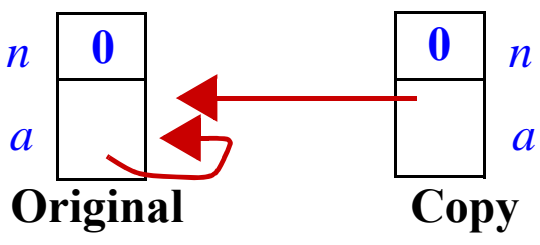

As a start towards the solution (7.5), note that in the history of the library criticism arose — independently of considerations of static verification — over the export status of *copy*, which originally was public by default. Imagine a class with attributes (fields) *a*, a reference, and *n*, an integer, and the property $(n = 0)$ => $(a =$ **Current**$)$. As illustrated, copying or cloning an object for which $n = 0$ yields an inconsistent new object since its *a* field points to the original, not the copy. There is no contradiction since the class is incorrect: one of its features, *copy*, does not preserve the invariant. But that feature is inherited, and it is not a good idea to force programmers to redefine it if they do not actually want to provide copying. A wiser solution is in *ANY* to export *copy* and *cloned* selectively to a class *COPIABLE*, from which a class must inherit if it is to provide copy and clone capabilities to its clients. With the O-rule, the call **Result**.*copy* (**Current**) only assumes invariant clauses involving properties of *COPIABLE*, and does not conflict with other invariant properties of a specific class. On exit, *copy* must yield the full invariant.

## 4.4 Monogamy

The next example comes from the dependent delegate paper [25]. It is delicate not only to verify but also to write in the first place, and provides a good benchmark for OO verification techniques.

We want a class *PERSON* with queries *spouse*: **detachable** *PERSON* and *is_married*: *BOOLEAN* satisfying the invariant property

*is_married* => ((*spouse* ≠ **Void**) **and** (*spouse*.*spouse* = **Current**))

The reason *spouse* is declared **detachable** is that not everyone is married, so **Void** has to be a valid value for *spouse*; it is in fact its initial value on creation of a *PERSON* object. To change this value by making a person married we need a routine *marry* (*other*: *PERSON*)[19]. Enforcing monogamy, the routine has preconditions **not** *is_married* and **not** *spouse*.*is_married*. (Also, *spouse* ≠ *other*.) Its job includes marrying the *other* object back to the current person; but *marry* cannot just call *other*.*marry* (**Current**) without causing infinite recursion. [25] presents several intermediate solutions, which it shows to be incorrect, and arrives at the following implementation for *marry*:

```
set_married                      -- 1 Body of marry; not final version, see 7.5.
other.set_married                -- 2 (numbers added for reference)
set_spouse (other)               -- 3
other.set_spouse (Current)       -- 4
```

with two utility routines:

```
set_married do is_married := True end
set_spouse (other: PERSON) do spouse := other end
```

Each does part of *marry*'s job: setting *is_married*, and setting the *spouse* reference. Here too the principles of OO design suggest exporting both routines selectively to *PERSON* itself (as in [25]).

18. The first reaction was to add instructions disabling invariant monitoring before the call to *copy* and restoring it afterwards, a kludge (but in some sense a precursor to Boogie's "unpack/pack"). Then the implementation of *cloned* was moved to a built-in function in the run-time system. With this article's O-rule is possible to restore the original code.

19. *other* is not detachable since the chosen spouse object must exist.

This example is a concentrate of problems that can occur with invariants. Put another way, it is the OO verification nightmare.

First, furtive access:

- At the start of the first qualified call *other*.*set_married*, the invariant (highlighted above) is trivially satisfied for *other*, since *is_married* is false for that object. But at the end it is violated since *is_married* is now true for *other* but its *spouse* has not been set yet.
- For the second qualified call, the invariant will be satisfied at the end, but violated at the start.

Next, reference leak. Assume as a thought experiment that *set_spouse* were exported instead of being just a utility routine for *marry*[20]. Then we could write

*Dominique*.*marry* (*Claude*)
*Dominique*.*set_spouse* (*Leslie*)

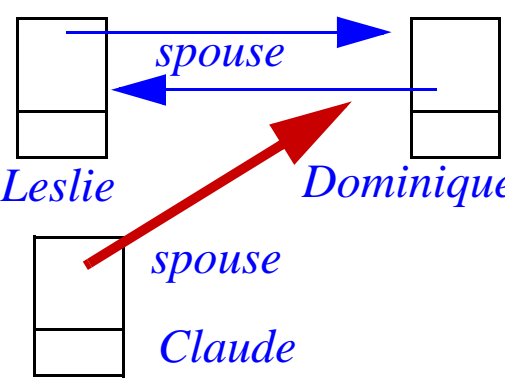

After the second instruction, *Dominique*.*spouse* is *Leslie*, but *Claude*.*spouse* is still *Dominique* from the first instruction, so *Claude*.*spouse*.*spouse* is *Leslie*, causing *Claude* to violate its invariant property *spouse*.*spouse* = **Current**.

We may brush off this form of the example since *set_spouse* should not be exported, precisely because it does not preserve the invariant. We cannot replace *set_spouse* by *marry* above, because *marry* has precondition clauses including **not** *is_married*. But consider a new routine:

*divorce* **do** *spouse* := **Void** ; *is_married* := **False** **ensure** *spouse* = **Void** ; **not** *is_married* **end**

This is not how we should normally write such a routine: if we divorce A from B, we should also divorce B from A. Nevertheless, the version given preserves the invariant — as would any routine that sets *is_married* to false — and satisfies its postcondition. The code

*Dominique*.*marry* (*Claude*)
*Dominique*.*divorce*

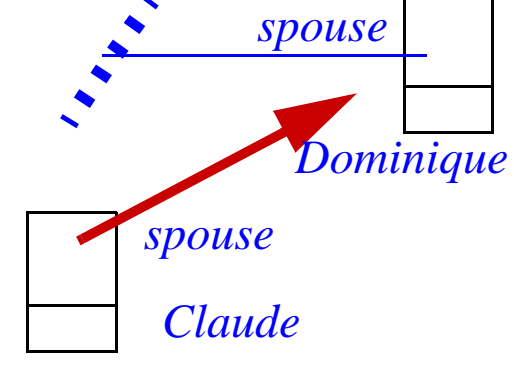

causes the *Claude* object to violate its invariant. The disturbing part, as in other cases of reference leak, is that we broke an object invariant[21] through an operation (*Dominique*.*divorce*) that does not even mention or involve that object.

It is a good bet that whoever can verify the monogamous marriage example has a shot at solving the OO program verification problem.

## 4.5 Linked lists and linkables

Consider the standard implementation of a linked list structure, where an object of type *LINKED_LIST* is exported to clients, with operations to insert, remove and access elements; the implementation uses objects of a type *LINKABLE*. The *LINKED_LIST* object contains general bookkeeping information about the list, such as the number of elements, *count*, as well as a reference *first*[22] to the first *LINKABLE* cell, and possibly references (not shown) to other such cells.

The *LINKABLE* cells are not meant for independent use: they are subservient to the *LINKED_LIST* header object. All other clients should go through that header. By having sole con-

20. Variables all of type *PERSON*, objects all initialized as needed.
21. Reminder: "object invariant" means the class invariant applied to one particular instance.
22. In the actual library this feature is called *first_element*. For brevity this article uses the name *first* (reserved in the library for a *public* feature returning the *value* of the first element, i.e. *first_element*.*item*).

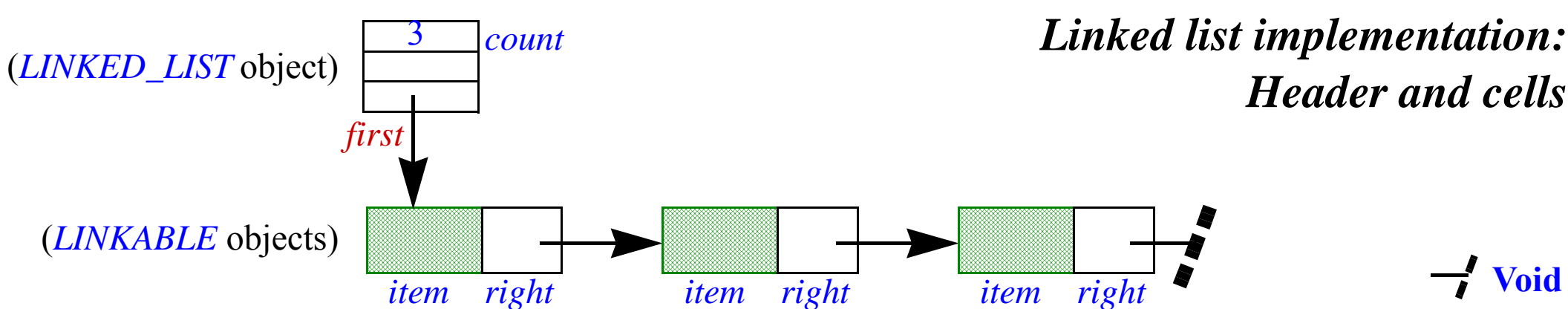

***Linked list implementation: Header and cells***

trol over list elements, the header can maintain crucial properties of the list, expressed in the class invariant of *LINKED_LIST*, which might contain such clauses as:

- *first*.*right_acyclic*, where *right_acyclic* expresses[23] that the sequence of cells obtained by repeatedly following *right* links has no cycle.
- *count* = *first*.*computed_count*, where *computed_count* counts the number of elements encountered by following *right* links as far as possible.

In the terminology of this article, these clauses make *LINKABLE inhibit LINKED_LIST*.

Reference leak in this example would occur if some object other than the list header somehow got hold of the reference *first*, and used it as the target of routine calls to modify cells directly, bypassing the header object's control. We can use the same model as in the unregistered observer example, using *LINKABLE* instead of *POINT* (the inhibiting class) and replacing both *UNREGISTERED_OBSERVER* and *MISCHIEF* (the inhibited classes) by *LINKED_LIST*.

Of course the classes in the actual library do not engage in any such mischievous games, but verification needs to ascertain this property rigorously. A simple way to mischief would be for *LINKED_LIST* to export the default version of *copy* and *cloned* (the non-redefined shallow-copy version, which simply copies an object field by field); then we could produce two list headers referring to the same *LINKABLE* objects:

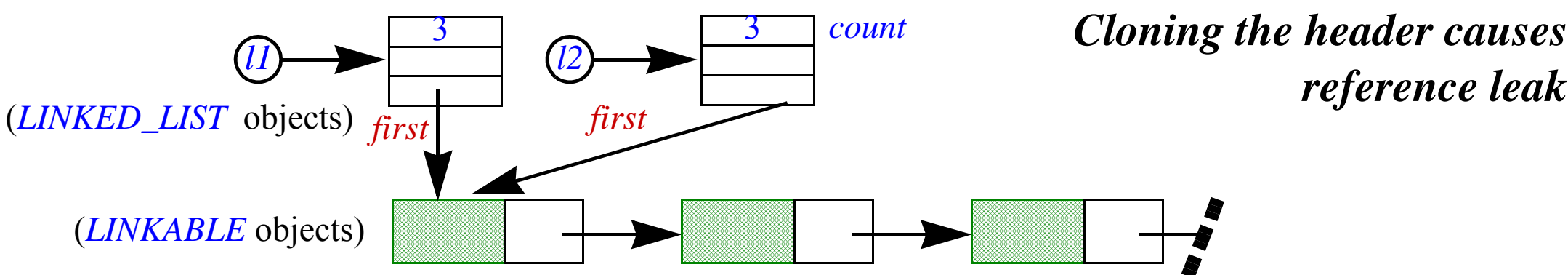

***Cloning the header causes reference leak***

Initially they are consistent, as shown, but it is easy, for example through a call *l2*.*remove_last*, which removes the last element, to invalidate *l1* — its associated list now has two elements, but its *count* field still says 3 — without any explicit operation on *l1*. Reference leak in its full horror.

The proper approach is to make the default copy and clone routines secret, and in *LINKED_LIST* export the redefined versions, which duplicate the *LINKABLE* list cells along with the list header. But methodological advice is not a substitute for verification. The sad reality in this example is that the default copy routine *will* preserve the invariant of *LINKED_LIST*, even though it introduces a time bomb into the data structure in the form of a reference leak.

## 4.6 Merging lists

The last example continues with linked lists, focusing on the tricky operation that merges a list with another: the routine *merge_right* (there is also *merge_left*) in class *LINKED_LIST*. It takes

23. Assuming the appropriate formalism, not explored further here.

another list as argument; as illustrated, the call *list1.merge_right* (*list2*) concatenates the second list to the first, and zeroes out that second list (making its *first* reference void).

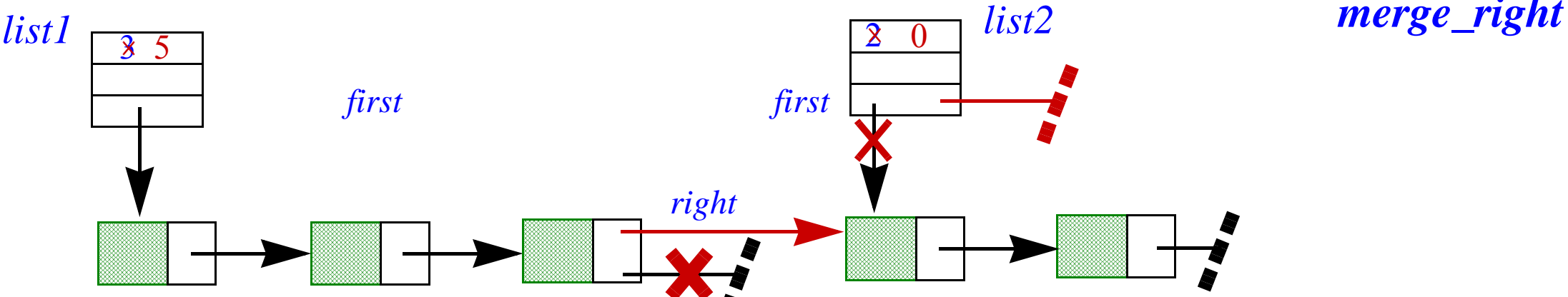

What makes things delicate is that programming the operation efficiently *requires* a reference leak:

- In implementing *merge_right* (*other*)[24], we could stay on the safe side of information hiding by taking the successive values *v* in the *other* list — not the actual *LINKABLE* cells, but the list values, obtained through the public interface of class *LINKED_LIST* — and adding each of them to the end of the first list through the call *extend* (*v*), then at the end use *other.wipe_out*, calling another public routine *wipe_out* to empty the list.
- This implementation, however, is O (*count2*), where *count2* is the number of elements in the second list, assuming we initially have access to the last cell of the first one. We can get O (1) by using the implementation illustrated in the last figure: directly link the last *LINKABLE* cell of the first list to the first cell of the second one, then set the second list's *first* to void.

We may call the two implementations "dumb" and "smart"[25]. The smart implementation has to use *other.first* directly; in other words, it must take advantage of a reference leak from the second list to the first. Perhaps this is the reason why *merge_right* turned out to be one of the most bug-prone routines in the entire library; for example an early version omitted the precondition clause *other* ≠ **Current**, without which things can turn messy.Verification, both static (AutoProof) and dynamic (AutoTest) uncovered the bug [38][26]. (A different version of *LINKED_LIST* is part of the EiffelBase 2 library, which was mechanically verified as part of Polikarpova's PhD work[40].)

Such routines should of course be programmed with particular attention, and presumably no error remains in the current version. But the problem is more general. The routine *merge_right* indulges in precisely the kind of internal pointer shenanigans, with the attendant reference leaks, that in other cases yields mayhem. This is the scary part: why is shallow-copy bad and *merge_right* OK? How do we distinguish — in a form that can be codified into simple formal rules, and taught to a verification system — the good cases from the bad?

Like monogamous marriage, list merging is a challenge and test of effectiveness for any OO verification method hoping to address class-invariant-related issues.

## 5 Misconceptions

The preceding analysis has identified the two serious problems raised by class invariants. It is also important to understand that some aspects gravely studied in the literature are *not* problematic.

24. *other* is *list2* in the example call.
25. "Smart" and "dumb" in terms of run-time performance. Verification gives us another perspective: the first implementation is obviously correct, the second one far trickier to verify, hence not so smart if correctness is the main concern. But in practice no project will renounce efficiency for the sake of verification. We need both.
26. The bug also served as a testbed for the AutoFix system for automatic suggestions of bug correction.

A common misconception is to believe that invariants govern all routines and all calls. In reality it only governs exported routines and qualified calls. But the Wikipedia entry on the topic [43] currently[27] proclaims that "*temporary breaking of class invariants between private method calls is possible, although not encouraged*"[28], without giving any reason why it should be discouraged[29].

No such reason exists; and breaking the invariant in unqualified calls is not only allowable but essential. The first item in the charter of the programmers' inalienable rights must surely be the right to abstraction: take any part of the code, give it a name, plus (optionally) arguments to make it parameterizable. Then you may replace the code, where it appeared, by a call to the resulting routine. To such a routine, the class invariant is irrelevant. We might for example rewrite the implementation of *move_left* (page 8) as:

```
go_back        -- This line is the only change; it previously read: x := x – 1
if x < 0 then x := 0 end
```

introducing a routine *go_back* which simply performs *x* := *x* – 1. This routine does not preserve the invariant. It will naturally be secret (private), since we must not allow qualified calls *p*.*go_back*. But unqualified calls *go_back* are fine, as here in *move_left*.

This misconception is bad enough in Wikipedia but we find it in scholarly articles as well. For example [4] has, in a class *T*, an exported routine *M* with an unqualified call to a routine *P*. It notes: "*at the time P is called, the object invariant*[30] *for the T object is not certain to hold*" and goes on to develop solutions to this imagined problem. But the invariant is irrelevant here: we simply have an issue of indirect recursion, which arises identically in a non-OO setting, and is susceptible to classical treatment in axiomatic semantics. Reference [15] has a similar example and discussion.

Surprisingly, all these articles cite OOSC 2, which explained the property emphatically[31]:

> *Qualified calls, of the form a.f (…), executed on behalf of a client, are the only ones that must always start from a state satisfying the invariant and leave a state satisfying the invariant; there is no such rule for unqualified calls of the form f (…), which are not directly executed by clients but only serve as auxiliary tools for carrying out the needs of qualified calls. As a consequence, the obligation to maintain the invariant applies only to the body of features that are exported either generally or selectively; a secret feature — one that is available to no client — is not affected by the invariant.*

*Dynamic* invariant verification in EiffelStudio naturally follows this policy: a qualified call triggers an invariant check, an unqualified call does not. This means in particular that a qualified call **Current**.*r* (*a*), using the current object (**Current**, *this*, *self*) as its target, is not exactly the same as the unqualified call *r* (*a*): one triggers the invariant, the other does not[32].

There is indeed a serious callback problem with invariants, causing furtive access and studied below (6.4); but it only arises for *qualified* calls.

---

27. As of September 2016.
28. It is better to correct a Wikipedia entry than to criticize it, but I leave this task to others.
29. This practice of guarding against a certain practice ("*X considered harmful*"), without explaining why or giving actionable criteria for when it is acceptable and when not, is unfortunately common in today's discussions of software methodology. Right or wrong, Dijkstra knew better.
30. Without any good reason, some authors use "*object invariant*" for what has for almost half a century been known as a class invariant. One of the reasons for rejecting this practice is that it makes it impossible to use "object invariant" to denote the class invariant *as applied to a particular object* (an instance of the class).

# 6 Towards a proof rule for object-oriented programming

The examples of section 4 give us a good measure of the difficulty of using invariants in the practice of verification. The first step is to develop a solution to furtive access (reference leak follows in section 9). While recent work has tended to propose "methodologies" for dealing with invariants, we seek a more solid result: a proof rule covering the semantics of object-oriented programming through its central mechanism, qualified calls.

To arrive at that result — the O-rule, already revealed in the introduction — we will go through several intuitively appealing initial attempts and show why they are unsound or insufficient. The final version appears in the next section (7).

## 6.1 First attempts

Let us start with the non-object-oriented world, where calls are all unqualified. The classical rule for such calls comes from another Hoare article [11]. We call it the N-rule, for Non-object; like all others in this discussion, it appears in a form that does not handle recursion (adding recursion, through techniques found in the literature, would make the rules a bit heavier, and is independent of the issues under examination). The notations are as in section 1.

/N/

$$\frac{\{Pre_r\,(f)\}\ body_r\ \{Post_r\,(f)\}}{\{Pre_r\,(a)\}\ \textbf{call}\ r\,(a)\ \{Post_r\,(a)\}}$$

This rule states that the effect of a call is the effect of executing the body after substitution of actual for formal arguments. It captures the fundamental role of routines (subprograms, methods): abstracting some computation by giving it a name and parameterizing it.

*Soundness*: yes (needs to be adapted for recursion).

*Usefulness*: not just for non-OO programming, but also for unqualified calls in an OO context.

We need to adapt the idea to object-oriented programming, where routine calls can be qualified, as in **call** $x.r\,(a)$. A first simple version is:

/O1/

$$\frac{\{Pre_r\,(f)\}\ body_r\ \{Post_r\,(f)\}}{\{x.Pre_r\,(a)\}\ \textbf{call}\ x.r\,(a)\ \{x.Post_r\,(a)\}}$$

(The final O-rule is actually /O6/, so we have some way to go.)

*Soundness*: yes, with same qualifications as N-rule.

*Usefulness*: this rule is simply the N-rule: although the syntax is object-oriented, the rule simply treats $r$ as if it had one more argument, $x$, enjoying a special syntax but no special semantics. It does in the formal world what compilers for OO languages, including compilers that generate C code, do: add an argument, representing the target, to every routine. But it does nothing to reflect the specific nature of OO programming, in particular the distinguished role of the target of qualified calls (corresponding to the notion of "current object" at any time during execution).

The class invariant is part of that specificity. To express that every routine available for qualified calls preserves the invariant, we may add the invariant to both the precondition and the postcondition, in both the hypothesis and the conclusion:

31. Citation from [23], 11.8, p. 370.

32. While there is little reason to write **Current**$.r\,(a)$ rather than the simpler form, the more relevant case is $x.r\,(a)$ where at execution time $x$ could sometimes denote the same object as **Current** and sometimes not.

/O2/

$$\frac{\{\, INV \wedge Pre_r(f)\}\ body_r\ \{\, INV \wedge Post_r(f)\}}{\{\, x.INV \wedge x.Pre_r(a)\}\ \mathbf{call}\ x.r(a)\ \{\, x.INV \wedge x.Post_r(a)\}}$$

For all rules on exported routines involving the invariant, there has to be a companion rule involving creation procedures, corresponding to the initial vertical transition in the Fundamental Picture (page 6). Creation does not assume the invariant on entry, but has to ensure it on exit. Using the name *make* for a typical creation procedure, the rule here is:

/C2/

$$\frac{\{\, DEF \wedge Pre_{make}(f)\}\ body_{make}\ \{\, INV \wedge Post_{make}(f)\}}{\{x.Pre_{make}(a)\}\ \mathbf{create}\ x.make(a)\ \{\, x.INV \wedge x.Post_{make}(a)\}}$$

where *DEF* expresses that all fields have the standard initialization values (such as 0 for integers).

*Soundness* (of /O2/ with its companion /C2/): yes, with same qualifications as N-rule.

*Usefulness:* /O2/ fails to capture part of the role of the invariant. It recognizes that the preconditions and postconditions of all exported routines (routines available for qualified calls) share a property, *INV*, but does not take advantage of the resulting preservation property, since every call must still ensure $x.INV$ on entry. So in fact it is still the N-rule: similar to /O1/, with the common pre- and postcondition elements factored out.

OOSC 1 and 2 use the following version, which does take advantage of the invariant:

/O3/

$$\frac{\{INV \wedge Pre_r(f)\}\ body_r\ \{INV \wedge Post_r(f)\}}{\{\, x.Pre_r(a)\,\}\ \mathbf{call}\ x.r(a)\ \{\, x.Post_r(a)\,\}}$$

The associated creation rule C3 is the same as /C2/ without the addition of $x.INV$ to the postcondition in the conclusion.

*Soundness:* yes, with same qualifications as N-rule.

*Usefulness:* this version recognizes the specificity of the class invariant as a preservation property. But it uses that property purely inside the class, to define class correctness (a class is correct if every exported routine, starting in a state satisfying the invariant and the precondition, yields a state satisfying the postcondition and the invariant). This means for example that client of the *ACCOUNT* class using a bank account object *a* cannot rely on the property that $a.balance = a.in_total - a.out_total$[33]. So we are still missing an important part of the concept.

## 6.2 The verification process

/O3/ is only an imperfect step towards the final rule, but we can use it to understand how such rules determine the process of verifying OO software. The focus is on static verification (proofs), but the discussion also has consequences (explored further in 7.7) on using invariants as dynamic checks for testing and debugging.

All versions of the rule have the same general form, differing only in the invariant-related properties that are added to the precondition and postcondition in the hypothesis (above the line), the conclusion (below the line) or both.

The proof process has two parts:

33. Since the lists *in* and *out* in the class *ACCOUNT* should most likely be secret (private to the class), we assume exported features *in_total* and *out_total* which return $in.total$ and $out.total$.

- Verifying a class — once and for all. This step applies to every routine $r$ of the class the hypothesis part of the rule.
- Once a class has been verified, verifying its clients. This step applies to every call $x.r\ (a)$ the conclusion part of the rule.

If the client relation is cyclic (two or more classes have qualified calls to each other), the steps are not as neatly distinct and the process becomes iterative. It is still useful to study them separately.

### *6.2.1 Proving the correctness of a class*

Proving the correctness of a class means establishing, for every routine $r$:

- If $r$ appears in unqualified calls, its N-rule correctness: $\{Pre_r\ (f)\}\ body_r\ \{Post_r\ (f)\}$.
- If $r$ is exported, to either all or some clients (details in section 7.2), hence available for qualified calls, its O-rule correctness, which in all versions of the rule has the form $\{INV' \wedge Pre_r\ (f)\}$ $body_r\ \{INV' \wedge Post_r\ (f)\}$ where $INV'$ is $INV$, as in /O3/, or some part of $INV$ in later versions.
- If $r$ is a creation procedure, its C-rule correctness (again similar for all versions): $\{DEF \wedge Pre_r\ (f)\}\ body_r\ \{INV' \wedge Post_r\ (f)\}$.

The conditions are not exclusive: a routine can be both usable internally in unqualified form and exported to clients; and it may be available both for calls and for creation[34]. In such a case, the routine's proof of correctness must include all the applicable rules.

The class proof process is modular: to establish the above properties, assuming the ancestors of a class have been verified, it suffices to examine its text. The second property states that the class text declares $r$ as exported, not that some qualified call (which could be anywhere in the entire system) actually uses $r$; and the third, that it declares $r$ as available for creation, not that some creation instruction (again anywhere) uses it.

Once a class — that is to say, each of its routines — has been proved correct in this way, we can prove properties of calls to these routines, typically to prove the correctness of the calling routines and classes.

### *6.2.2 Reasoning about unqualified calls*

For an unqualified call $r\ (a)$, the N-rule applies: we must establish that $Pre_r\ (a)$ holds before the call (obligation) and may deduce that $Post_r\ (a)$ holds after the call (benefit).

### *6.2.3 Reasoning about creation instructions*

For a creation instruction **create** $x.make\ (a)$ (which creates a new object with default fields, then updates them by applying the creation procedure $make$[35]), the C-rule applies: we must establish $Pre_{make}\ (a)$ before the call and may deduce $INV' \wedge Post_{make}\ (a)$ afterwards.

### *6.2.4 Reasoning about qualified calls*

To reason about a qualified call $x.r\ (a)$, we use the conclusion part of the rule (below the line):

- We *have to establish on entry* the precondition part of the conclusion, which always includes $x.Pre_r\ (a)$ (to verify a call unless we must know that the routine's precondition initially holds for the target object). In /O3/ this is all, but /O2/ also forces us to establish $x.INV$ which, as

34. C++ and, following it, Java and C# use a special convention for creation procedures: they are not features of the class but use the overloaded class name. In these languages the third case is disjoint from the first two.

35. The simplified version **create** $x$ is a shorthand for **create** $x.default_create$ using the class's version of the default creation procedure.

noted, loses the benefit of the very idea of invariant: if it is an invariant, it should have been maintained by previous calls; it should not be our job as client to establish it again.

- We *may deduce on exit* the postcondition part of the conclusion, which always includes $x.Post_r(a)$ but should also (as in and subsequent versions beginning with /O4/) $x.INV$, so that we know the call leaves the object in a sane state.

#### *6.2.5 Benefits and obligations*

In comparing the various forms of the rule, we should remember the usual give-and-take of Design by Contract [21, 23]. If we add elements to the left part $P$ of a Hoare-style property $\{P\}$ $A$ $\{Q\}$:

- In the hypothesis part of a rule, it makes the process of proving class correctness easier since we can assume more.
- In the conclusion part, it makes the process of reasoning about a call harder since we have to establish more.

If we add elements to the right part $Q$:

- Proving class correctness is harder.
- Reasoning about a call is more rewarding since we can deduce more.

Comparing the conclusion parts of /O2/ and /O3/ in this light (their hypothesis parts are the same) shows that /O2/ is both better and worse: better because it gives us the target invariant $x.INV$ on exit, but worse because we have to establish that same property on entry. The rule of the game, as we resume our effort to get the right version of the rule, is to keep the benefit and get rid of the undue obligation.

## 6.3 The ideal rule

To treat the invariant as a conservation property, so that we may assume it on exit but do not have to establish it on entry, the desirable rule would be:

/O4/
$$\frac{\{INV \wedge Pre_r(f)\}\ body_r\ \{INV \wedge Post_r(f)\}}{\{\ x.Pre_r(a)\ \}\ \textbf{call}\ x.r(a)\ \{\ x.INV\ \wedge x.Post_r(a)\}}$$

The associated creation rule C4 is the same as C3.

/O4/ directly reflects the Fundamental Picture; we will call it the "*ideal rule*". In a simple world it would be final.

*Soundness*: only in the absence of qualified callbacks (see next), reference leaks and recursion.

*Usefulness*: when applicable, the ideal rule captures the essence of object-oriented programming and makes reasoning about OO programs simple and effective.

The ideal rule will make a comeback in 7.6 as the "O'-rule", applicable when there is demonstrably no possibility of a callback.

## 6.4 Qualified callbacks

The problem with the ideal rule /O4/ is the risk — discussed as the "Dependent Delegate Dilemma" in [25] — of furtive access arising from qualified callbacks, which endanger the beautiful simplicity of the Fundamental Picture:

Argument lists have been omitted. Routine $r$ gets called on target $x$ and start executing on the attached "Object S". It makes a qualified call of target $y$, attached to "Object T". The routine in that call is $u$. It execution in turn makes a qualified call whose target $z$ happens — tough luck! — to be attached to S. That call uses the routine $v$; its execution terminates, execution of $u$ on T resumes and terminates; control returns to the rest of the execution of $r$ on S. The problem is that the callback $z.v$ catches S in a temporary state, marked by a red dot in the figure, where the invariant has no reason to hold. Unlike the unqualified callbacks of section 5, this is a qualified call, for which we would normally expect the invariant to apply.

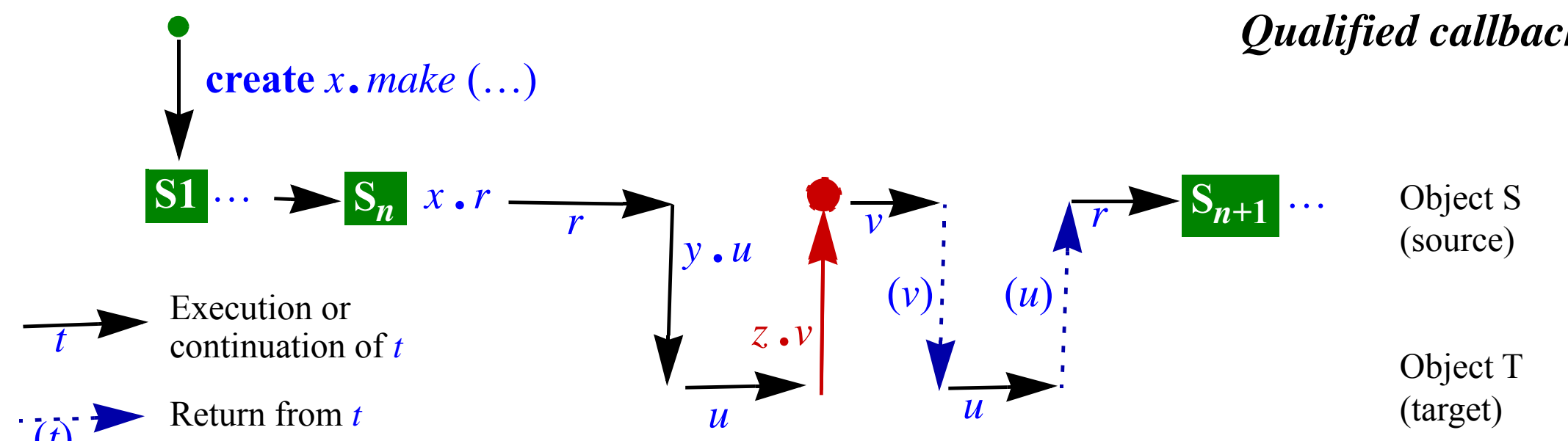

The scenario shown is not just a theoretical possibility but arises with normal program schemes, as shown next. But before we start adding another Rube Goldberg contraption to the programming language, two observations are in order:

- The invariant violation *does not matter* for the original routine $r$. From $r$'s perspective, the call $y.u$ and its consequences such as $z.v$ are just steps in the algorithm, and need not concern themselves with the invariant of S's class any more than unqualified calls do.
- The reasoning behind rule /O4/, however, was that we can drop the $x.INV$ part on the left of the conclusion line (as it appeared in /O2/, added to $x.Pre_r(a)$) because qualified calls occur in sequence, each one finding a sane state[36] and leaving a sane state. But here such a call happens in the middle of another, destroying this reasoning.

The first of the preceding observations led to a solution described in the Dependent Delegates paper [25]: treat qualified callbacks like *unqualified* calls. An exported routine must satisfy some version of the O-rule ([25] uses /O3/); if it can be used in a qualified callback, that solution requires it *in addition* to satisfy the /N/-rule (the non-OO version that does not involve the invariant, and is applicable to unqualified calls).

This solution does not use the right version of the O-rule, and making it modular requires extra work. Still, it has the merit of simplicity and is a step in the right direction.

## 6.5 The strongest rule

To be safe we may simply require that whenever computation branches out of an object, as $r$ does in the last figure, the object clams up — makes sure its invariant holds — to be ready for any eventuality. At home you can dress, or not, as you like, but before getting out you make sure you have something on. In the kitchen metaphor from section 3.2, you are having a small office party and have messed up the kitchen, and need to get out to ask your boss a question; but first you clean the place, just in case during the discussion he decides that he needs a cup of coffee.

36. Reminder: a sane state is one satisfying the invariant.

The "Boogie methodology" [3, 4, 15, 18, 33, 39] has a ghost instruction[37] *wrap* functioning as an assertion — to be verified by a prover — that, at the given program point, the invariant holds. The clamming-up idea is similar. The difference is that it is not an instruction that programmers must write, but instead, it will be part of the proof rule. Also, we do not need a counterpart to Boogie's *unwrap* instruction, which states that the invariant might not be satisfied. In general, the Boogie methodology understands a class invariant *INV* not in the classical sense dating back to Hoare and OOSC — a property that holds on entry and exit of qualified calls — but as a shortcut for something like *is_wrapped => INV*. This view is disturbing. "*The kitchen must be clean between uses*" is a simple and clear rule. If every such rule automatically includes the implicit qualification "*unless otherwise noted*", the benefit of having any rules at all becomes doubtful.

The next version of the O-rule restores the classical view:

/O5/
$$\frac{\{INV \wedge Pre_r(f)\}\ body_r\ \{INV \wedge Post_r(f)\}}{\{\ INV \wedge x.Pre_r(a)\ \}\ \textbf{call}\ x.r(a)\ \{\ x.INV\ \wedge x.Post_r(a)\}}$$

(Not the final rule yet, but we are getting close.) The invariant part added to the postcondition is *x.INV,* as before; but on the precondition side it is just *INV*. (Some readers may find it more clearly expressed as **Current**.*INV*, or, Java-style, this.*INV*.) We do not want *x.INV* here: that would just be /O2/ (which, as we saw, loses the value of the invariant since the client has to establish it before every qualified call, whereas we should be able to trust that the previous operations on our target object have preserved it). Adding *INV* on the client side expresses the clamming-up obligation: before getting out of the house, we put something on.

As far as I know, no one has proposed any such rule; while it requires one more improvement to be applicable in practice, it captures the interplay, fundamental to an understanding of object-oriented programming, between the client and the supplier, reflected in their invariants.

In the usual get-and-take of Design by Contract, the concept of invariant brings the client both an extra obligation and an extra benefit: before a qualified call, you must ascertain, in addition to the precondition, your own invariant; after the call, you are entitled to know, in addition to the postcondition, that the supplier object satisfies its invariant.

As with all rules, these observations apply to both static and dynamic verification:

- With a static prover such as AutoProof/Boogie, "*you*" means the prover, "*ascertain*" means obligation to prove (*assert* ghost instruction in ESC-Java, JML, Boogie etc.), and "*entitled to know*" means that the prover may add the property to its list of established assertions (*assume*).
- With run-time contract monitoring as in EiffelStudio, "*you*" means the contract monitoring mechanism, "*ascertain*" means evaluate (raising an exception if the clause evaluates to false, doing nothing more otherwise), and "*entitled to know*" means not having to evaluate anything.

We will now refine the rule to avoid clamming up objects more than strictly necessary.

## 7 The O-rule

Designing programming support for verification is a trade-off between three criteria: soundness; flexibility (how few forms of expression we have to *renounce*); and ease of use (how few verification-oriented annotations we have to *add*). The Boogie methodology does well on the first two,

37. "Ghost" in the sense that it only serves for verification and has no influence on correct executions. Some publications use the names *pack* and *unpack* instead of *wrap* and *unwrap*.

but explicit wrapping and unwrapping — and we have not even seen ownership yet — removes the prospect of "Verification As a Matter Of Course", usable by ordinary programmers.

Rule /O5/ fails the flexibility test. Requiring that an object always satisfy the full class invariant before branching out is too much. Many practical examples do not meet this requirement.To turn /O5/ into a more realistic rule we will take advantage of the notion of restricted export.

### 7.1 About furtive access

All the section 4 examples causing furtive access involve restricted exports:

- In the observer pattern (4.2), *OBSERVER* and *POINT* export their mutually relevant features, *subject* and *observer*, to each other only.
- In the attempt to define cloning from copying (4.3), we noted the proposed policy of exporting *copy* selectively to a class *COPIABLE*.
- In monogamous marriage (4.5), *PERSON* exports the utility routines *set_spouse* and *set_married* to itself only.

The decision to use restricted exports in each of theses cases, although not indispensable (the features could have been public, arises from good design methodology. But the examples illustrate that in cases of qualified callbacks, and furtive access in general, it is natural to give the corresponding features a limited export status. That cannot be a coincidence!

This observation will yield the final form of the O-rule.

### 7.2 About selective exports

As a reminder, the selective export mechanism extends the Information Hiding principle [37] by expressing that not all clients are created equal. Most OO languages, as noted, provide some form of it (C++ friends, .NET assembly). The effect is to restrict the availability of certain operations of a class to specified classes (hence "friend"). When one of the feature clauses of class *C* reads

```
feature {A, B, C}
    r1 … Routine declaration …            -- The example ignores arguments
    r2 … Routine declaration …
```

it specifies (as part of the static type rules) that a qualified call *x*.*r*1 or *x*.*r*2 is only permitted (for *x* of type based on *C*[38]) if it appears in *A*, *B*, *C* or one of their descendants. A clause **feature** without further qualification introduces fully exported features and is equivalent to **feature** {*ANY*}; secret features are introduced in a **feature** {*NONE*} clause[39].

Export restrictions also govern to the use of a class as its own client: for *x* of type *C*, *x*.*r1* is only permitted if we did list *C* as above, even though the **feature** clause appears in the text of *C* itself. Unqualified calls such as a plain **call** *r1* (…) are, of course, always valid within *C*, but qualified calls have to abide by the normal client export rules[40].

What is the relation of this concept to furtive access? We noticed earlier that qualified callbacks are similar in spirit to unqualified calls. A more precise version of this observation is that they

38. Every type is "*based on a class*". Often the class and the type are the same thing but the "*based on*" concept accounts for generic types such as *LIST* [*INTEGER*] as *LIST* [*PERSON*]: in both cases the "*base class*" is *LIST* independently of the generic parameter. The base class determines the applicable operations (features).

39. In the lattice-like multiple inheritance graph of classes, *ANY* is the top, ancestor to all classes, and *NONE* is the bottom, descendant to all classes.

40. Some OO languages muddle this matter, from a lack of attention to the difference between qualified and unqualified calls.

usually come from friends. (In the middle of the office party gone a bit wild, you may not want the boss to enter the kitchen, but there is nothing wrong in letting your buddies in.)

With the final O-rule and its companion, the Export Consistency rule, this advice ("*they usually come* …") will becomes an obligation: we will only accept callbacks from friends. "Friend" being an informal term, we need a finer analysis.

## 7.3 Slicing an invariant according to privilege

Consider a routine with a certain export status: for example *r*1 above is exported to *A*, *B* and *C* (and their descendants). Also consider an invariant clause involving features of the same class:

*is_ready* => (*balance* > 0 **and** *other*.*credit* > 0)

(An invariant, like other assertions, is made of any number of such clauses, implicitly "and"-ed.) The clause involves three features of the enclosing class: *is_ready*, *balance* and *other*. These features also have an export status. $INV_r$, for any such feature *r* (for example $INV_{other}$), will denote the part of the invariant including only those clauses with no more export rights than *r*.

**Definition**: for a feature *r* in a class *C*, $INV_r$ is the invariant of *C* deprived of any clause that contains an unqualified call to a feature of *C* exported to classes to which *r* is not exported.

In deciding which clauses to retain, we only consider features of *unqualified* calls. In the example the export status of *credit*, in its own class, does not matter, since *credit* is the feature of a *qualified* call. But the target of that call, *other*, does matter (its use is an unqualified call).

The clause above will be part of $INV_r$ (where *r* is a feature of the enclosing class) if and only if *other*, *is_ready* and *balance* are only exported to the same classes as *r* or a subset of them.

For callbacks emanating directly or indirectly from a routine *r*, considering $INV_r$ rather than the whole *INV* reflects the need for fine granularity. To let your boss in, you want a clean kitchen. To let the safety inspector in, you want no chairs blocking exits and no cables lying on the floor. To let your coffee-loving colleague in, you want the coffee machine ready. In each of these cases, the other two conditions may not be relevant; the applicable condition is tailored to the selected client. For the world at large (arbitrary clients), all invariant properties must hold.

Hence the final version of the O-rule, differing from /O5/ in the replacement of *INV* by $INV_r$ in the precondition part:

/O6/

$$\frac{\{\, INV_r \;\wedge Pre_r\,(f)\;\}\; body_r\; \{INV\;\wedge Post_r\,(f\,)\}}{\{\, INV_r \;\wedge x.Pre_r\,(a)\,\}\;\textbf{call}\; x.r\,(a)\;\{x.INV \wedge x.Post_r\,(a)\}}$$

For a fully exported feature *r*, $INV_r$ is the same as *INV*, so /O6/ reduces to /O5/.

The soundness of the O-rule requires a simple consistency condition preventing *x*.*r* from making a qualified call *y*.*u* that would affect properties beyond those accessible to *r*. Section 8 will present this condition, the Export Consistency rule.

## 7.4 Invariants on entry and exit

Both the hypothesis and conclusion of the O-rule (/O6/) add a different version of the invariant to the precondition and to the postcondition: $INV_r$ on the left, *INV* (*x*.*INV* in the conclusion) on the right. This asymmetry is surprising at first but inevitable. On entry, we have to take into account the possibility that the call may be furtive, catching the object in an intermediate state:

- In the observer case (4.2), at the time of the call *observer*.*update*, an observer that has not yet made itself consistent with its subject.
- In the cloning case (4.3), at the time of the call **Result**.*copy* (**Current**), a freshly created object that has not yet filled its field with sane values.
- In the marriage case (4.4), at the time of the call *other*.*set_spouse* (**Current**), a person object that has done only half of what it takes to get married.

The use of $INV_r$ on the precondition side addresses the issue on entry. On exit, however, we cannot resort to the same technique. The problem is that once a qualified call terminates, we have no way to know which qualified call, if any, will hit the same target next. It might be another consequence of the current higher-level call to $r$, in which case $INV_r$ would be fine, but it might be a completely independent call targeting the same object at any later time — including, in a concurrent object, from a different processor. In such cases we have no guarantee that $INV_r$ will suffice; we have no better bet than the full $INV$.

In other words, we may know that a given call comes from a friend, but since we do not know that the next call will we always have to leave the object in a state suitable for any client.

The unpleasant consequence is that when proving the hypothesis $INV_r \wedge Pre_r(f)\ body_r\ \{INV \wedge Post_r(f)\}$ to prove correctness of the class (6.2.1) we must establish the full invariant $INV$ on exit but can only assume the partial invariant $INV_r$ on entry. How do we know that the implementation of $r$ will ensure properties beyond $r$'s own export privilege?

In many practical cases, this requirement raises no difficulty. The observer case is typical. The invariant (4.2) is

*faithful*: *x* = *subject*.*x*
*backlink*: *subject*.*observer* = **Current**

and the body of *update* has the postcondition *x* = *subject*.*x*. Since *update* and *subject* have the same export privilege, being exported selectively to *POINT*, while *x* (in *OBSERVER*) is public, the invariant slice $INV_{update}$ includes the clause *backlink* but not *faithful*. As a consequence:

- Since *update* does not affect *subject*, it conserves the *backlink* property[41].
- Its postcondition is exactly the same property as *faithful*.
- As a consequence, *update* yields the full $INV$ starting from $INV_{update}$.

Such cases are common, but in others we will need to establish that a qualified call to a routine $r$ leaves the target object in a state satisfying not only $INV_r$ (as part of the routine's normal business) but also $\overline{INV_r}$, using this notation to denote the remaining clauses of the invariant. To this effect, we may simply add $\overline{INV_r}$ to the precondition $Pre_r$ of $r$, and check that the implementation of $r$ does not affect properties with a higher export privilege than $r$, hence does not invalidate $\overline{INV_r}$. In practice this policy means transferring part of the responsibility to clients: on entry to a call $x.r(a)$ we must establish, as part of $x.Pre_r(a)$ (and in addition to $INV_r$ on the source object) that $x.\overline{INV_r}$ holds. We are partly back to the naïve rule /O2/ which forced us to establish the invariant on the target object before a qualified call — but only for a part of that invariant, not the full $x.INV$.

### 7.5 Addressing furtive access examples

The O-rule immediately legitimates two of the preceding furtive access examples:

41. The absence of any change to *subject* is a "frame property". It should either appear explicitly in the postcondition of *update* or be inferred automatically [13].

- In the observer case, as just seen, the call *observer.update* is now correct: it does not require the invariant clause *x = subject.x* on entry, but guarantees it on exit, while preserving the other clause and hence ensuring the full *x = subject.INV*.
- In the cloning case, if *copy* is selectively exported, the call **Result**.*copy* (**Current**) does not need the rest of **Result**'s object invariant, but ensures it on exit (as *copy* must do).

The marriage case is more subtle. In 4.4 the invariant was

*is_married* => ((*spouse* ≠ **Void**) **and** (*spouse.spouse* = **Current**))

and the implementation of *marry*, taken from [25]:

```
set_married                  -- 1
other.set_married            -- 2
set_spouse (other)           -- 3
other.set_spouse (Current)   -- 4
```

with the auxiliary routines *set_married* **do** *is_married* := **True end** and *set_spouse* (*other: PERSON*) **do** *spouse* := *other* **end**. These routines are exported only to *PERSON* itself, so we again have no problem on entry (remember that the export status of a routine *r* of a class *C* governs all qualified calls to *r*, so that even in *C* itself *x.r* for *x* of type *C* is only valid if *r* is exported to *C*). But the call *other.set_married* does not yield *other.INV* since on exit *other.is_married* is true but *other.spouse* still void. In other words, this code from [25] is incorrect in the formal framework of the present article as expressed by the O-rule.

To rectify the situation it suffices to reverse lines 2 and 4. In fact as long as 4 appears before 2 the relative timing of the other instructions does not matter.

How bad was the original error? If the code is exactly as given, *set_spouse* and *set_married* do no more than their job of setting a field, and the computation is sequential, it does not matter. But if these conditions do not hold, mischief can occur:

- In a concurrent setting, a different thread could access *other* after instruction 2, expecting the invariant to hold and hence its *spouse* field to be non-void, causing a null-pointer dereferencing if it tries to access it[42].
- Even without concurrency, it is easy for such code to go wrong. Let us tweak the example by adding public features *is_minor* and *drink*. Both *drink* and *marry* have the precondition **not** *is_minor*, and *marry* also has **not** *other.is_minor*. The class has a new invariant clause: *is_married* => **not** *is_minor*. We insert in *marry*, after instruction 2, instructions *drink* and *other.drink* — starting to celebrate even before the marriage is finalized. Now assume a bizarre implementation in which *set_married* sets the age to a value satisfying *is_minor*, and *set_spouse* restores its original value. Bizarre, but in principle correct since the routine overall preserves the invariant. Since instruction 1 now invalidates the invariant, we know that we should replace the call *drink* by **if not** *is_minor* **then** *drink* **end**. The instruction *other.drink* is different: it is a qualified call to a public routine, and hence should be able on entry to assume that *other* satisfies it object invariant, which implies the routine's precondition; but at this point (after instruction 2 and before instruction 4) the invariant is broken. With its precondition not satisfied, *other.drink* can malfunction; we are allowing a minor to drink.

42. This scenario is not possible in Eiffel: first, the concurrency rules [30] ensure that *marry* will have exclusive access to *other* during its execution; then, the void safety mechanism will reject any code that could cause a null-pointer dereferencing [27]. But other concurrency mechanisms and frameworks may not have those guards.

These examples — which the O-rule rejects — illustrate the preceding explanation (7.4) of why a qualified call $x.r$ should *always* yield a state satisfying the full target invariant $x.INV$.

### 7.6 The no-callback O'-rule

In the conclusion of the O-rule, the reason for adding $INV_r$ to the precondition in the conclusion part is to protect against qualified callbacks; but in practice they rarely occur. If we have the guarantee that $r$ will not produce any, the O-rule simplifies to a variant of the earlier "ideal rule" /O4/:

/O'/
$$\frac{\{INV_r \wedge Pre_r(f)\}\ body_r\ \{INV \wedge Post_r(f)\}}{\{x.Pre_r(a)\}\ \textbf{call}\ x.r(a)\ \{x.INV \wedge x.Post_r(a)\}}$$

This version (O'-rule) facilitates the proof of correctness of qualified calls, since in the conclusion we have less to establish before allowing a call: just the precondition, no invariant or any slice of it. Without the risk of qualified callbacks there is no more need to clam up. (If you go see your boss during the wild office party and are sure the meeting will all take place in his office, no need to clean the kitchen first.)

How realistic is it to expect a no-callback guarantee in a qualified call? The answer has two parts:

- In a general setting, establishing the absence of callbacks requires full *alias analysis*: we have to find out whether any variable anywhere in the code could hold a reference to the source object). The "alias calculus" [29, 13] is an effort to provide an automated mechanism for alias analysis. The implementation, however, is not yet fully operational. In addition, alias analysis is not naturally modular.
- In specific cases, it may be possible to obtain the guarantee more simply. An example is a routine that has no qualified calls whatsoever, although this is an implementation (rather than specification) property. Another is the case of a call to a routine $r$ of a pre-existing library, which cannot possibly call into newer application code. It is not always easy in practice to guarantee such a property, since in some cases — particularly event loops and various forms of UI programming — the routine of the callback is known not statically but through a variable (closure/agent/delegate/function pointer, see [36]).

So while we might long for the more straightforward O'-rule, the O-rule is our default tool.

Two more consequences of the fundamental proof rule are worth noting.

### 7.7 Dynamic checking policy

Class invariants and other assertions can serve not only for proofs (static verification) but also as dynamic checks that can be enabled at run-time for testing and debugging purposes. Proofs are almost always better than run-time checks, but they are not always possible and are not yet part of mainstream development processes. In the practical application of Design by Contract techniques, run-time assertion monitoring remains an essential tool, dramatically facilitating testing and debugging.

Run-time contract monitoring, when activated, evaluates the invariant after creation, then both before and after every qualified call. It was always known that in an ideal world — more precisely, a world satisfying the Ideal Picture (3.2) — the "before" check would be superfluous; but also that the problems discussed in this article, furtive access and reference leak, may cause external interference between the last check of an object's invariant and the next qualified call on that object. OOSC includes a detailed discussion of the both-before-and-after policy as a response to these problems[43].

In light of the O-rule, the "before" check seems unnecessary: we need at most to check $INV_r$ for the source object, not the target's invariant. But unless we also address the risk of reference leak (see below), it is not yet time to perform this simplification.

### 7.8 Selective exports and invariants

An already given citation from OOSC states[44]:

> *The obligation to maintain the invariant applies only to the body of features that are exported either generally* ***or selectively****…*

No deep thinking was probably involved: selectively exported features simply seemed to fall in the same category as exported ones. But surely there is a gradation between a fully exported feature, formally understood as declared in **feature** {*ANY*}, and a fully secret one, **feature** {*NONE*}. Why consider all intermediate cases, such as **feature** {*B*, *C*}, equivalent to the first?

Only with the present discussion does the answer appear: invariant semantics follows (in the pre-part) a parallel gradation, obtained by considering every clause of an invariant individually.

## 8 The Export Consistency rule

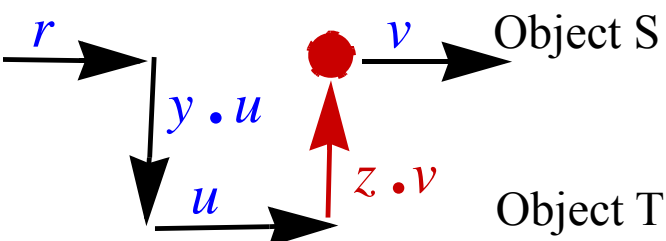

To apply the final O-rule (/O6/) soundly, we must enforce proper information hiding. Assume that *x*.*r* (ignoring arguments) is executed on behalf of an object S and then, as in the figure illustrating callbacks (6.4, page 22, see extract on the right), calls back into S, for example through a call *z*.*v* where *z* happens to be attached to S. If the routine *v* has broader export privileges than *r*, it could modify properties of the class that appear in the invariant outside of $INV_r$. This scenario is incompatible with the soundness of the O-rule.

It is in fact — regardless of verification concerns — incompatible with the principle of information hiding. In a class *PRIVATE* consider a routine *r* exported to *FRIEND* but not to *FOE*. Class *FOE* may not call *p*.*r* (*args*) for *p* of type *PRIVATE*; but it can easily bypass that restriction without inheriting from *PRIVATE* and without any modification to *FRIEND* or *FOE*. Just add a simple class

```
class SPY inherit FRIEND feature
    bypass (p: PRIVATE; args: …) do p.r (args) end      -- args declared like formals of r.
end
```

Then have *FOE* inherit from *SPY* and use *bypass* (*p*, *args*). (If you dislike using inheritance for such purposes, just write **create** *sp*; *sp*.*bypass* (*p*, *args*) for a local variable *sp* of type *SPY*.)

Surprisingly, no one seems to have complained about this information hiding loophole so far, but it should be corrected. The appropriate language rule requires the following property[45].

**Definition**: a routine *r* satisfies the **Export Consistency rule** if no routine of a qualified call appearing in the text of *r*, or of a routine called by *r* unqualified, has a greater privilege than *r* or, if *r* is a redefinition, its precursor.

---

43. [23], 11.14, page 410.
44. [23], 11.8, page 370, emphasis added.
45. Web browsers such as Firefox offer a "private" mode protecting users from sites' tracking. From a public window, you may choose to open a link in a new public or private window; but from a private window, the new window can only be private. This is the same idea as the Export Consistency rule.

The "privilege" of a routine (short for "qualified call privilege") is, informally, the set of routines that it can use in qualified calls. Here are the formal details expressing this intuition:

- The privilege of a class is a partial function from classes to routines[46] of those classes. An example is {[*A*, {*r*, *s*}], [*B*, {*t*}]},[47] indicating permission to access *r* and *s* from *A* and *t* from *B*.
- The order relation between privileges is simply the subset relation (between partial functions: $p \leq q$ if every [argument, result] pair of *p* is also in *q*).
- A privilege contains explicit elements directly induced by feature clauses; if *A* has the clause **feature** {*A*} *r*, *s* (feature declarations omitted) and *B* has **feature** {*A*} *t*, then the privilege of *A* includes the example function above.
- In addition, the privilege of a class *A* contains implicit elements: the function pair [*X*, *u*] for every public (fully exported) routine *u* of any class *X*[48], as well as the privileges of all the ancestors of *A* (since exporting to a class means also exporting to its descendants).
- The privilege of any routine of a class *A* is the privilege of *A*.
- The last part of the Export Consistency rule prohibits any redefinition (overriding) of a routine from making qualified calls that the rule would prohibit for its precursor (the original version).

The notion of privilege covers the classic information hiding rule[49], which states (in traditional terms) that one may use *x*.*s*, in a routine *r* of a class *C* with *x* of type *T*, only if *T* exports *s* to *C*: just rephrase it as "the privilege of *r* must include the pair [*T*, *s*]". The Export Consistency rule goes further by preventing *r* from calling (in qualified form) a routine that would circumvent this restriction. While it is necessary for the soundness of the O-rule, it makes sense independently, and will be proposed as an addition to the language standard [7].

As a verification rule, Export Consistency is modular. To apply the rule it suffices, when compiling a routine *s* or analyzing it for verification, to compute its privilege — as compilers must do anyway, to enforce standard information hiding — and include it in the interface information for the routine. Then the processing of any routine *r* that includes a qualified call *y*.*s* should check that the privilege of *s* is no greater than the privilege of *r* and, if applicable, its precursor.

The privilege includes only information on the routine and the interface of some of the routines it calls. "Some of" because one may choose to hide part of that information, for example the internal routines used by a library routine; clients will simply not be able to call them directly. Such hiding is compatible with the rule since it makes the published privilege *smaller*.

"Implicit elements" mentioned in the definition of "privilege" serve conceptual purposes only and need not be computed. The tools should only compute explicit elements, from selective export clauses of the form **feature** {*X*, *Y*, …}. Then in checking the validity of *y*.*s* in the text of *r*:

- If *r* is public, the call satisfies the rule regardless of *s*'s privilege.
- If *r* is selectively exported and *s* public, the call violates the rule regardless of *r*'s privilege.
- If *r* and *s* are selectively exported, *s*'s privilege must be no greater than *r*'s privilege.

Each case only uses privilege information about *r* and *s*. The rule requires no global information.

46. More generally, features/members (including attributes/fields).
47. Using braces {…} for sets and brackets […] for pairs and representing a partial function as a set of pairs; here the first element of each pair is a class and the second element a set of routines.
48. Considering implicit elements is convenient but not conceptually necessary: since a public routine is formally equivalent to one declared in an explicit **feature** {*ANY*} clause, implicit elements follow from the other cases.
49. In the Ecma standard [7], it appears in the definition of "available for call" in clause 8.7.13 and the rest of 8.7.

The following property summarizes this article's solution to furtive access[50]:

**Proposition**: if *x* is of a type *T* that has been proved correct (6.2.1) and *r* satisfies the Export Consistency rule, any call $x.r\ (a)$ executed under $INV_r \ \wedge x.Pre_r\ (a)$ will yield $x.INV \wedge x.Post_r\ (a)$. (In other words, the Fundamental Picture holds, even in the presence of qualified callbacks.)

# 9 Reference leak and the inhibition rule

We will now develop a solution of the second problem, reference leak.

## 9.1 Reference leak examples: a quick reminder

Reference leak, as we saw, can occur in the marriage case (4.4): the code

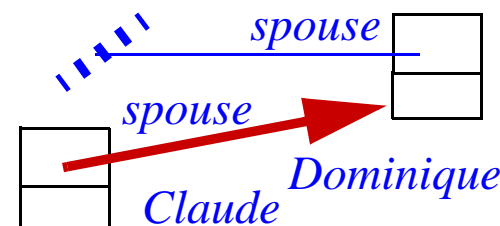

```
Dominique.marry (Claude)
Dominique.divorce
```

causes the *Claude* object to violate its invariant. The reference leak scheme that we witnessed for unregistered observers (4.1) can also arise with standard observers (4.2) if they have the extra invariant property $subject.x < L$:

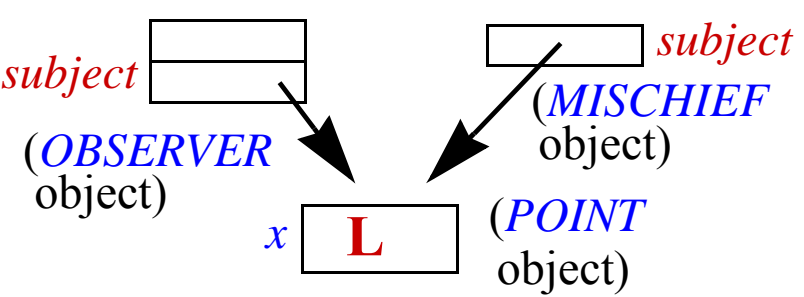

```
class MISCHIEF create make feature
    subject: POINT
    obs: OBSERVER
    make do create subject ; create obs.set (subject) end
    mess_up do subject.move_left end
end
```

So can the "divorcing" scheme: in class *POINT*, a procedure *remove_observer* that sets *observer* to void will preserve the invariant

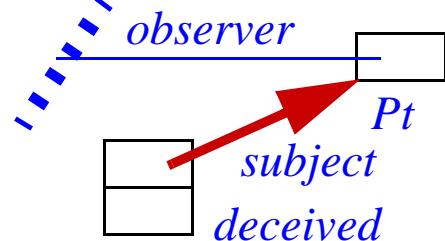

```
(observer ≠ Void) => (observer.subject = Current)
```

but then executing the instructions

```
Pt.set_observer (deceived)    -- deceived could be Current.
Pt.remove_observer
```

will break the invariant of *deceived*.

Note, however, that here since *POINT* exports *set_observer* only to *OBSERVER*, this code has to appear in *OBSERVER* or a descendant. This observation will help us towards a solution that is not applicable in the marriage case, where *spouse* is public.

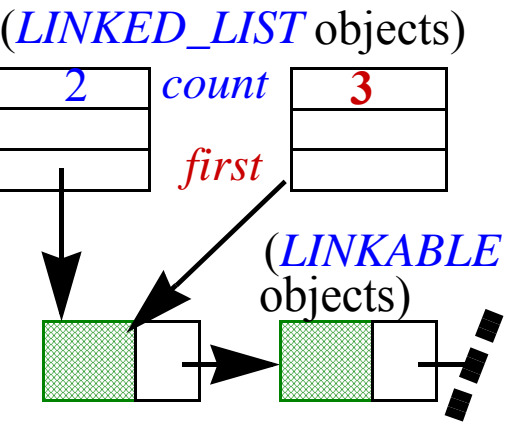

The linked list examples of 4.5 and 4.6 involved corrupting the *first* field of a linked list object (the object that represents the list header) so that it will point to the wrong *LINKABLE*, such as one from another list, endangering the sanity of the header object since — for example — the *count* field might not give the actual number of list elements. In the example, there were initially three elements, and procedure *remove_last* was applied to the first list object, which has its *count* correctly updated to 2, but the second object still has 3.

50. "Proposition" because the property, justified by a detailed discussion in the previous sections, is more than a conjecture, but — not having been proved mathematically — less than a theorem.

Finally, remember that even the basic example of bank accounts with their lists *in* and *out* of deposit and withdrawal operations will suffer from if one of the lists, *in* in the figure, leaks to another object C, which through its own operations can invalidate the account object's invariant *balance = in.total – out.total*.

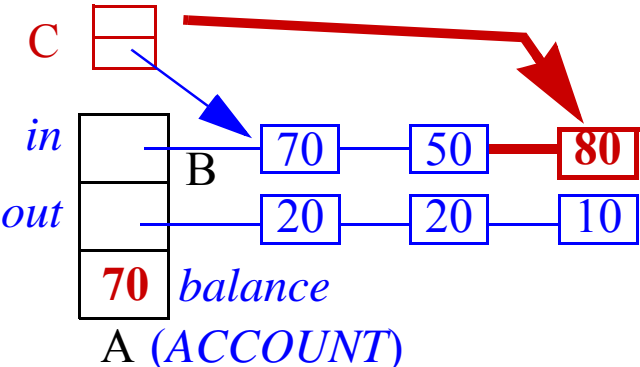

## 9.2 Reference leak conditions

Going beyond individual examples, we must define the general pattern that cause reference leak. As correctly identified in [3], reference leak happens because of invariant clauses containing qualified calls *q.r* where *q*, of some type *T*, is a query of the enclosing class and *r* is a query of *T*. If *q* is detachable (possibly void), a common form is

*q* ≠ **Void** => *q.r* -- In this case *r* is a boolean property.

Terminology: any occurrence of *q.r* with *q* of type *T* in the invariant of class *D* causes *T* to **inhibit** *D* through *q* **with** *r*. *T* is the **inhibitor**, *I* the **inhibited** class, *q* the **inhibiting tag**, and *r* the **inhibiting query**. The concepts transpose from classes to the corresponding objects: in the figure on the right, object B, an instance of *T*, inhibits object A, an instance of *D*, through *q*.

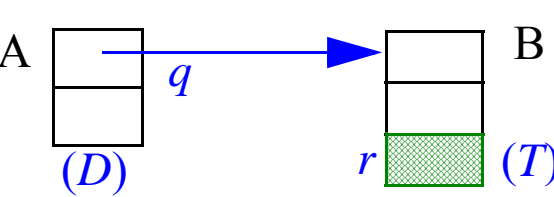

The reference leak scenario is simply that a third object C obtains a reference *q1* to *B* (the reference has been "leaked" to C) and uses it to modify the value of *r*, through some operation *q1.mess_up*. In the examples:

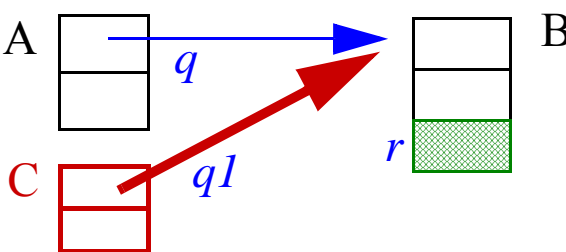

- The banking-operations-list class inhibits *ACCOUNT* through *in* and *out* with *total*.
- *POINT* inhibits *UNREGISTERED_OBSERVER* and *OBSERVER* through *subject* with *x*, and *OBSERVER* through *subject* with *observer*.
- *OBSERVER* inhibits *POINT* through *observer* with *subject*.
- *PERSON* inhibits itself through *spouse* with *spouse*.
- *LINKABLE* inhibits *LINKED_LIST* through *first*.

The inhibition concept resembles the widely used notion of *ownership*[35][51], which in such situations would require the programmer to declare (in some extension of the language and type system) that A "owns" B. Ownership is too coarse-grained: it applies to objects in their entirety, whereas with inhibition the relation between A and B only applies to a specific tag *q*. Various objects may inhibit each other in different ways for different tags *q* and queries *r*.

In fact, while ownership is inherently non-symmetric (if A owns B, B cannot own A), inhibition can be symmetric. As just seen, *OBSERVER* inhibits *OBJECT* through *observer*, and *OBJECT* inhibits *OBSERVER* through *subject*. The relation, between classes, can even be reflexive. as with *PERSON* inhibiting itself through *spouse*.

A general solution to the reference leak problem can only be of two kinds:

- A condition under which reference leak will not occur. The inhibition rule, coming next, falls into this category.
- A condition under which reference leak does not invalidate the inhibiting object's invariant. The concept of tribe (section 10) is a tentative step in this direction.

51. The literature on ownership is huge; this citation is to one of the first publications, with no intended slight to authors of others.

## 9.3 External sanity

The first step towards a solution is, as elsewhere in this article, to remove layers of complication and realize that the problem may be less difficult than it seems: standard OO information hiding principles already go a long way. Selective exports played a major role in the solution to furtive access; they will be just as essential to addressing reference leak.

Any well-written implementation of an inhibition scheme will use restricted exports to limit access to features that could cause leak issues. For example *LINKED_LIST* will not export *first* and other features giving access to *LINKABLE* cells. Class *LINKABLE* will export its own features to *LINKED_LIST* only:

```
class LINKABLE [G] feature {LINKED_LIST}
    right: detachable LINKABLE [G]                      -- Next cell.
    item: G                                             -- Value stored in current cell.
    put_right (other: detachable LINKABLE [G])          -- Link to other.
            do … ensure right = other end
    put (value: G)                                      -- Set cell's value to value.
            do … ensure item = other end
end
```

Similarly, *OBSERVER* and *POINT* each exports the features that can cause reference leaks, respectively *subject* and *observer*, to the other class. (In the elementary example of 4.1 *UNREGISTERED_OBSERVER* does not follow this rule, but a carefully written version will.)

This common-sense policy gives us the first part of the inhibition rule:

**Definition**: Class *B* inhibiting *A* with[52] *r* satisfies the **external sanity clause** if it selectively every feature that may modify the value of *r*, and *r* itself if it is of a reference type, to *A* only.

Determining which features may modify *r* is a simple modular check, using only the text of class *A*[53]. (We could even use a stronger version of the rule, trivial to implement and possibly good enough in practice: apply the export restriction to any feature of *A* that has an argument or result of a type that conforms to the type of *r*.)

If *r* is of a reference type, exporting it is prohibited, since a call *r*.*some_operation* could modify the associated object. For a non-reference type such as *INTEGER* (also called an "expanded" or "value" type, exporting *r* is harmless.

## 9.4 Internal sanity

The external sanity clause almost rules out reference leaks, although in this "almost" lies thirty years of verification research. The clause tells us that harmful leaks only matter for objects of type *A*. Objects of other types can have all the leaked references they like to *B* objects (*LINKABLE* cells in the example), and we do not care: they will not be able to use them for mischief since they cannot apply any operations to them.

All that remains is the case of other *A* objects having their own leaked references to *B* objects that are supposedly under the control of the initial *A* object — their gateway. This case could indeed occur, for example with the following two routines added to *LINKED_LIST* [*G*]:

52. Remember that an invariant clause containing *q*.*r* causes inhibition "through" *q* and "with" *r*.

53. The rule also applies to descendants of *B*, but inheritance does not threaten modular verification here since descendants may only restrict privileges, not broaden them.

```
leak_and_mess_up (thief: LINKED_LIST [G]) do thief.mess_up (first) end
mess_up (f: LINKABLE [G]) do f.put_right (Void) end
```

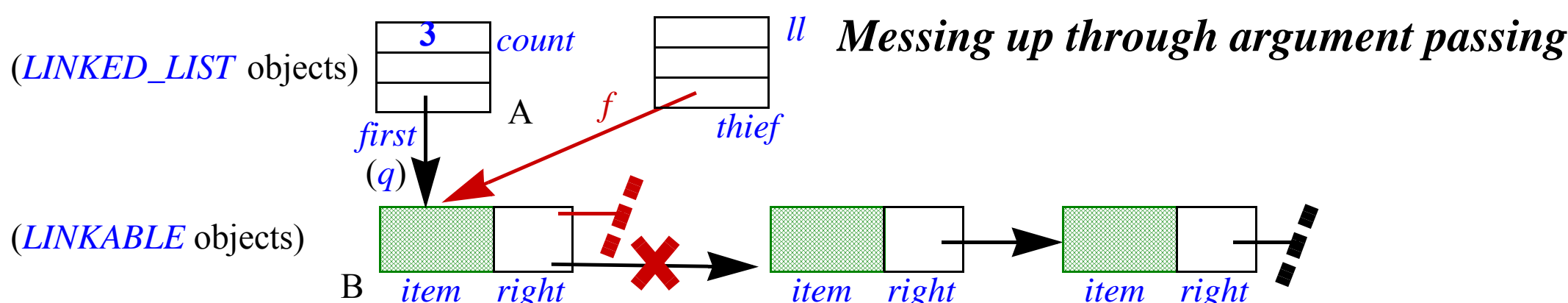

***Messing up through argument passing***

A call to *leak_and_mess_up* (*ll*), where *ll* is another linked list, will, as shown, transform any non-empty list into a one-element list, invalidating the *count* field and the corresponding invariant.[54]

Here is the analysis. If *B* inhibits *A* through *q*, the external sanity clause takes care of leaks to would-be thieves of types other than *A*; but we must also keep *A* objects, other than the original inhibited object A[55], from messing up with the inhibitor object B, known to the original though *q* (*first* in the example). Such a "thief" object could access B through its own leaked reference *f*, and use it to perform *f*.*modify*. The routine *modify* (*put_right* in the example) is available to the thief since it is exported to the class *A*; but we do not want any object other than A to use it on B.

How can the thief obtain such a leaked alias *f* of A's *q*? Regardless of who — A, the thief or a third party — created the inhibitor B, the reference to B had to be passed to at least one of the two A objects through a qualified call *x*.*mess_up* (…) where *x* (denoting either A or the thief) is of type A, and the routine *mess_up* either:

- Has, as in the example, a formal argument *f* whose value could be a reference to B.
- Is a function whose result could be a reference to B.

Expressing these conditions exactly would require global analysis on the object structure. The following stronger condition, however, is easy for a compiler to enforce as a simple addition to the existing rules of type checking and information hiding:

**Definition**: Class *B* inhibiting *A* through *q* satisfies the **internal sanity clause** if no feature of *A* other than *q* having an argument or result of a type conforming to *B* is available to *A* for calls or creation.

A "conforming" type is the itself or a descendant (taking genericity into account [23, 7][56]). A feature is "available" to a class if it is exported to it. This discussion assumes the Export Consistency rule (section 8), so that thieves cannot use tricks to gain access indirectly to the banned features.

**Definition**: a class inhibiting another through a query satisfies the **inhibition rule** if it satisfies both the external and internal sanity clauses.

**Proposition**[57]: an inhibition satisfying the inhibition rule cannot invalidate the inhibited class's invariant through reference leak.

---

54. The reader will have noted how disturbingly close this example is to the case, in principle legitimate, of *merge_right*. More on this point below.
55. It is generally not a good idea to distinguish things by font and color alone, but here it should cause no confusion that *A* and *B* are classes and objects A and B instances of each.
56. A more formal version of the rule should state the type constraint in terms not of *B* but of the type *T* of *q*; *T* is "based on" class *B*. See footnote 38.
57. As for the other fundamental "proposition" (see footnote 50), not yet a theorem but more than a conjecture.

This proposition is the second main result of the present article, addressing the second ope problem of class invariants, reference leak.

Lest us revisit the reference leak examples to see how they fare under the inhibition rule.

### 9.5 Non-leaking banking records

For bank accounts, external sanity means that in the list-of-banking-operations class the features that may affect *total* must be exported to *ACCOUNT* only. Since they include all routines that modify the list, we cannot directly use a list class from the library, for example declaring *in* and *out* as *LINKED_LIST* [*ACCOUNT*]. A simple solution is to use delegation, going through a class

```
class OPERATIONS_LIST feature {ACCOUNT}
    list: LINKED_LIST [ACCOUNT]
    … List features applied to list, for example count defined as list.count and so on …
end
```

For internal sanity, it suffices to make sure that class *ACCOUNT* declares no entity (routine argument, attribute, function result) of type *OPERATIONS_LIST*, except for *in* and *out*.

### 9.6 Non-leaking unregistered observers

For *POINT* inhibiting *UNREGISTERED_OBSERVER* through *subject* with *x* (4.1, 9.1):

- The classes as written do not satisfy the external inhibition rule since *POINT* publicly exports the routine *move_left*, which modifies *x*. To correct this problem, *POINT* must make *move_left* selectively exported to *UNREGISTERED_OBSERVER* only. There is no restriction on exporting *x* itself since it is of an expanded (non-reference) type, *INTEGER*.
- The classes also do not satisfy the internal rule, since *UNREGISTERED_OBSERVER* exports *set* publicly and hence to itself, making *MISCHIEF*'s leaking call *obs*.*set* (*subject*) possible. *UNREGISTERED_OBSERVER* has to exclude itself from the availability of *set*. It should only be available to classes such as *MISCHIEF* which create *UNREGISTERED_OBSERVER* objects and need to initialize them with a reference *subject* to the point they will be watching.

These changes limit the use of class *POINT*, making the status of its instances similar to those of *LINKABLE* cells in linked lists, which clients can only modify through a gateway object (the list header). Here the gateway is the *UNREGISTERED_OBSERVER* object.

Such a restriction becomes inevitable — per the inhibition rule — as soon as instances of one type, here *UNREGISTERED_OBSERVER*, rely for their sanity on the properties of instances of another, here *POINT*. Then we have to prevent instances of other types from breaking that special relationship: the penalty is that they may only access the relevant features (here routines modifying the *x* of a point) by going though an instance of the gateway type. (Section 10 points to a completely different approach, which would not require such hiding of critical features.)

Unlike ownership, the special relationship does not apply to the classes *in toto*, only to specific operations, here those modifying *x*. *UNREGISTERED_OBSERVER* does not "own" *POINT*; it is simply inhibited by *POINT* with *x* through *subject*. Instances of other classes can access any other properties of a *POINT* object without interference from *UNREGISTERED_OBSERVER*.

### 9.7 Non-leaking observers

For standard observers (4.2, 9.1), the first step, because of the inhibition of *OBSERVER* by *POINT* though *subject* with *x* (here in the invariant clause $x = subject.x$) is, as with unregistered observ-ers, to make sure that *POINT* exports *move_left* — and in a more general version of the class, any routine that can modify *x* — to *OBSERVER* only.

For other inhibiting queries, the Observer scheme as written almost passes the inhibition rule:

- Both classes satisfy the external sanity clause since *subject* and *observer* are the inhibiting queries involved in the respective inhibiting clauses, *make* and *set_observer* are the only routines that may modify them, and each of the two classes exports these features to the other one only (or, in the case of *make*, to no class at all).
- They almost satisfy the internal sanity clause, since they do not export these features to themselves for calls.

"Almost", because *OBSERVER* exports make *for creation*. Excerpting from 4.2:

```
class OBSERVER create make feature
    x: INTEGER
    …
feature {NONE}
    make (p: POINT) do subject := p ; subject.set_observer (Current) end
… See 4.2 for rest of class …
```

The definition of internal sanity (9.4) specifies (using the names of the classes and features of this example) that: "no feature of *OBSERVER* other than *subject* having an argument or result of a type conforming to *POINT* is available to *OBSERVER* for calls **or creation**". Having *make* generally available for creation, including to *OBSERVER* itself, violates this rule. Indeed the following code in *OBSERVER* would cause a reference leak:

```
thief: OBSERVER
create thief.make (subject)
```

making the current observer and *thief* share a subject point[58], and breaking the current observer's *backlink* invariant clause *subject*.*observer* = **Current**. So the problem is real.

To remove this problem, it suffices to satisfy the internal sanity clause by replacing the plain **create** clause with **create** {*OF*}, restricting observer creation to a class *OF* (for "Observer Factory") and its descendants, none of which is *OBSERVER* itself. (Alternatively, it could be useful to have syntax, not available today, for making features available to all classes *except* a designated one — or simply. for the sake of internal sanity, to all classes except the current one).

It is one of the tangible results of this article to allow the verification of a plainly written Observer pattern, without any verification-oriented annotation or other extra code:

- Even though the pattern relies on furtive access, accessing an object in a state that does not satisfy its invariant, the O-rule guarantees correctness.
- The inhibition rule shows that no reference leak can occur.

## 9.8 Non-leaking spouses

The marriage example resembles the Observer pattern in its use of mutual inhibition. But here, rather than two classes, we have only one, *PERSON*. The inhibition rule cannot apply, since the external clause would require *PERSON* to export spouse to itself and the internal clause not to.

58. Of course in the actual Observer pattern several observers can observe the same subject (see footnote 12). They still do not *share* the subject, only the *list* of subjects, to which the scheme described here exactly applies.

This impossibility is not an artifact of the rule but conceptually inevitable. If persons may know about other persons (as permitted if the internal clause does not apply) the earlier leaking scheme is possible from within class *PERSON*:

*Dominique*.*marry* (*Claude*)
*Dominique*.*divorce*

satisfying all the invariants of the objects involved. (As noted, this scheme assumes a bad version of *divorce* which does not divorce the other person; but there is no way to force the good version since the invariant to be preserved is that of another object, not named in the call. The bad version does preserve the invariant of the current object.)

An immediate solution is to remove *marry* from the public interface of *PERSON* and hand over marriage rights to a class *BROKER*, with attributes *spouse1* and *spouse2* of type *PERSON*. Routines *marry* and *divorce*, in their public versions, are now features of that class, used in the style *my_broker*.*marry* (*Dominique*, *Claude*). In a relationship reminiscent of that of *LINKABLE* to *LINKED_LIST*, class *PERSON* exports its marriage-related features such as *set_spouse*, *set_married* and (since it is a reference) *spouse* to *BROKER* only.

The effect of this solution on software design is significant, but any solution that both relies on a single class *PERSON* as described in previous sections and allows clients to use *marry* and *divorce* directly on *PERSON* objects will raises the possibility of reference leaks. The only other way is to take a different view of class invariants, as outlined in section 10.

## 9.9 Linked lists and linkables

With the inhibition rule we can make linked lists — temporarily setting aside *merge_right* — demonstrably leak-free. The external sanity clause already holds since *LINKABLE* exports its features to *LINKED_LIST* only. To satisfy the internal clause, it suffices to find the few features of *LINKED_LIST* taking arguments or yielding a result of *LINKED_LIST* type and remove *LINKED_LIST* from the set of classes to which they are exported.

In the actual library class, these features are all restrictively exported already, indicating that the class authors realized the need to handle them with care. More precisely, *LINKED_LIST* exports feature *first*[59] to *LINKED_LIST_ITERATION_CURSOR* (a class used to provide iteration capabilities on lists), respecting the strict internal sanity clause; but it also exports *first* to itself.

After removal of the export of *first* to *LINKED_LIST*, everything still compiles with the only exception— surprise! — of two routines: *merge_right* and *merge_left*.

The "dumb" version, not directly using references to *LINKABLE* objects, will of course compile.

## 9.10 Merging lists

What would it take to legitimize the "smart" versions of *merge_right* and *merge_left*? With the inhibition rule, it is not possible to write these routines in *LINKED_LIST*. We may use a solution similar to the introduction of a *BROKER* in the marriage case: use calls such as *h*.*merge_right* (*list1*, *list2*) using as the type of *h* a new class *LIST_HANDLER* to which *LINKED_LIST* and *LINKABLE* selectively export the appropriate features.

59. As noted (footnote 22), the actual name is *first_element*.

Since that approach again affects the software's design, we should again ask whether the inhibition rule's rejection of the smart *merge_right* is justified. Indeed, the rejection is not due to a quirk of the rule. It comes from a good reason: these routines could cause havoc. Compare two earlier figures, the first showing a harmful leak:

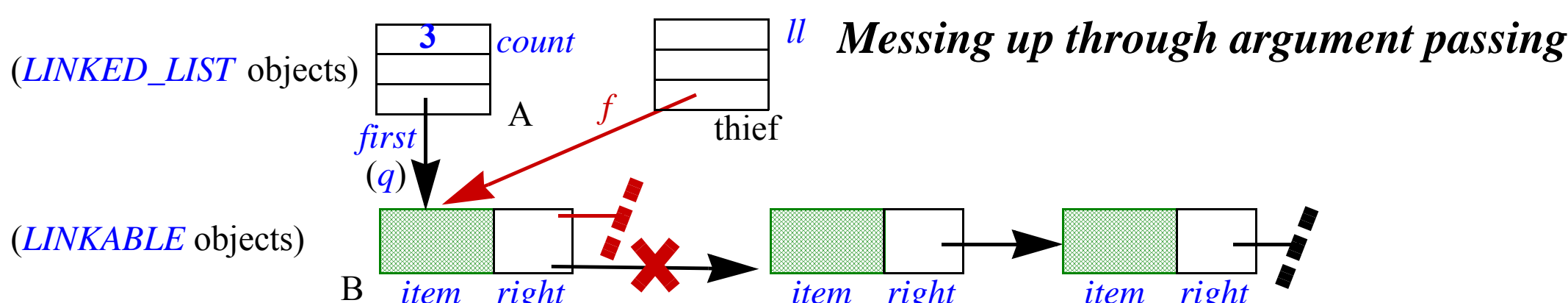

***Messing up through argument passing***

and the second one how *merge_right* works:

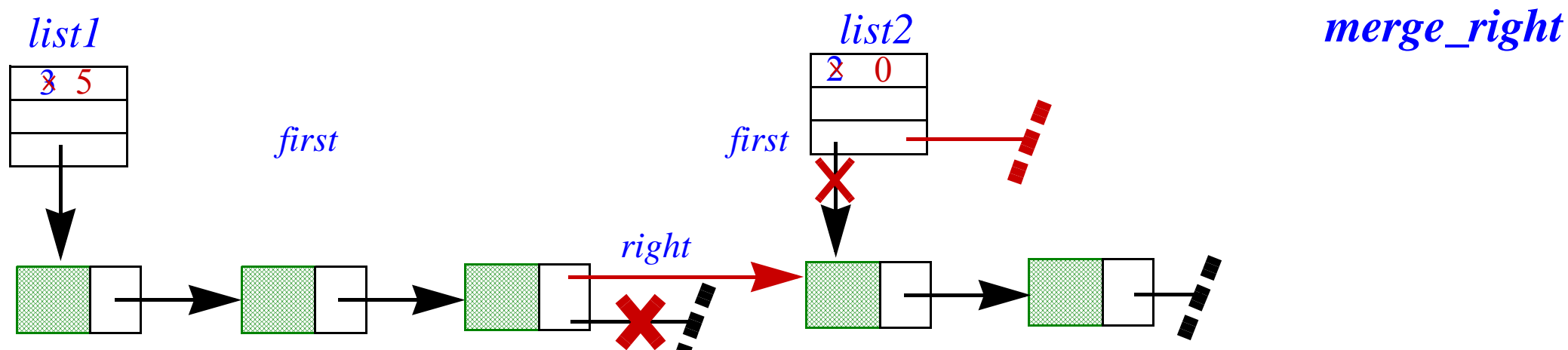

***merge_right***

They are remarkably similar. In both cases a list object gets access to the internals (*LINKABLE* references) of another object of the same type — as prohibited by the inhibition rule — and uses this access to reorganize the list cells; for evil in the first case, for good in the second, but it appears impossible to distinguish on the basis of a single class text.

It is for example easy to write a slight variation on *merge_right*, say *share*, through which *list1*.*share* (*list2*) will cause *list2* to refer to a part of the first list:

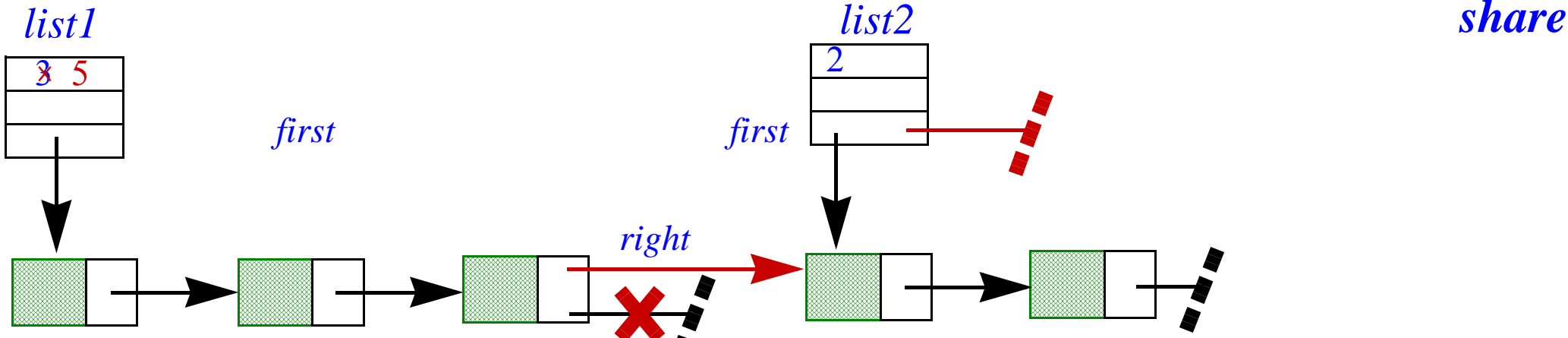

***share***

Such sharing contradicts the expectation that linked lists should never share the actual list cells. That expectation cannot, however, be made part of the invariant of *LINKED_LIST*: it is not a property of any particular list, but of the object structure as a whole. Even with the best invariants, expressing for example that the *LINKABLE* structure is acyclic, it is possible to write *share* so that it will preserve the object invariants of both *list1* and *list2*. The separateness property is also an invariant, but its scope is bigger than individual lists since it talks about the disjointness of *pairs* of lists. It can still be expressed as a class invariant, but not for the list class; a *LIST_HANDLER* class can host it.

## 10 Object tribes

This last part of the discussion sketches out a different approach. Unlike the previous sections, it is not fully developed; rather, it presents a possible direction for development, which, if deemed attractive, will require further work.

The one-way inhibition rule may seem to be the last word on the reference leak problem: we have seen strong arguments (although not a proof) that it is sound, that is to say sufficient to guarantee the absence of reference leaks; and the discussion of example violations shows it to be necessary in typical situations (as soon as we allow arguments or results of routines exported to *A* we can create a reference leak). Can there be room for anything else?

There can if we note that reference leaks are only harmful if they cause missing the leaked object's invariant during an update to another object. If every object's invariant included the invariant of other objects which it affects, then we would not need any particular restriction: normal invariant checking would suffice. For example, assume the invariant of every married person somehow includes the invariant of the spouse, and even if the person changes *spouse* that property remains in effect for the previous spouse (alimony). Then in the example

*Dominique*.*marry* (*Claude*)
*Dominique*.*divorce*

the first instruction adds *Claude* to the set of objects whose invariants *Dominique* follows: *Claude* and *Dominique* now belong to the same *tribe*. What defines a tribe ("all for one, one for all") is that the invariant of any object in the tribe includes the invariant of all others. The second instruction must preserve this combined invariant: then the sloppy version of *divorce*, which only sets *spouse* to void for the target object (here *Dominique*), will not pass verification; but the good version, which also sets it for the *spouse* object if any, will.

**Definition**: a set of objects is a **tribe** if they all have the same object invariant.

To make this definition practical, we must distinguish, for an object *o* in a tribe *T*, between its "written invariant" *I* (*o*), deduced from the class text (the class invariant as applied to *o*) and its effective invariant, which must be

$\forall\, o': T \mid I\,(o')$

For this discussion, tribes are useful in connection with inhibition. We may always assume that the set of objects is partitioned into tribes (if only through the trivial partition where every object makes up its own tribe); so we may talk of "the tribe of an object". From the tribe condition follows a new approach to reference leaks:

**Definition**: a set of classes satisfies the **tribe rule** if any of their instances that inhibits another, or inhibited it earlier in the execution, belongs to the same tribe.

(Note the reference to the past: in the current state of this work it is an open problem under what terms we may remove an object from a tribe.)

The tribe rule provides a completely different solution to the reference leak problem. If class *PERSON* satisfies the tribe rule, the first call above, *Dominique*.*marry* (*Claude*), will cause the *Claude* object to join *Dominique*'s tribe; the effective invariant, which includes both objects' invariants, then rules out foul play. We can handle all the other reference-leak examples in the same way, forcing into a single tribe: a subject and its observer; a linked list and the *LINKABLE*s on which it depends; two persons being married; two linked lists being right-merged (ensuring that each will be consistent, although not that they will be disjoint).

How could we make that approach work? A potential solution (described here tentatively and in general terms) would combine language and library properties. A library class *INHIBITED* provide features about an object's tribe: *tribe_has* (*x*) says whether *x* belongs to the current object's tribe, and *tribe_put* (*x*) adds *x* to that tribe. (Internally, they may be implemented as *tribe*.*has* (*x*) and *tribe*.*put* (*x*), where the attribute *tribe* is of type *TRIBE*, a class offering set-like operations.) The class *INHIBITED* has the invariant

$\forall\ o \mid (tribe_has\ (o) \wedge (o \neq \textbf{Current})) => o.INV$

or, in strict programming language notation

**across** *tribe* **as** *o* **all** (*o*.*item* /= **Current**) **implies** *o*.*INV* **end**

where *INV* yields an object's written invariant. The language rule is that any class *A* inhibited by a class *B* through a query *q* must:

- Be a descendant of *INHIBITED*.
- Have the invariant clause *tribe_has* (*q*) (qualified by $q \neq$ **Void** if the inhibiting clause is).

The second condition guarantees that the inhibitor object is in the tribe. In practice, maintaining it as part of the invariant will mean (as enforced by the verification process) that any routine which can change *q* includes an instruction *tribe_put* (*q*) or equivalent. For example, *marry* should now include *tribe_put* (*spouse*).

The tribe idea resembles the requirement to record "observers"[60] in the semantic collaboration approach [38, 40], but it is more general. "Observer" constructs in semantic collaboration use "ghost" variables, present only for verification and without a run-time effect. It is not clear whether tribe properties could have the same status; they are treated above as ordinary features, so they can serve for dynamic verification — run-time contract monitoring — as well as proofs. In this approach, every inhibited object has one more field, a reference to its tribe; in our examples, every *PERSON* will have a *tribe* field, and so will every *LINKED_LIST* (but not *LINKABLE*s, of which there are many more). The tribes themselves will, of course, use up space.

Beyond this particular language-plus-library solution (only outlined here, and possibly kludgy), the question to be addressed further is whether the underlying approach, the tribe rule, can lead to effective solutions to the reference leak problem, complementing the more straightforward solution of enforcing the inhibition rule.

## 11 Limitations

A formal proof of soundness is necessary, but is currently only in an early stage.

The approach has not yet been implemented in a compiler or verification system.

The internal sanity clause (9.4) could be made more flexible though alias analysis [13] (by considering not all entities of the inhibiting query's type, but only those which risk having the same value as that query).

The rules ignore recursion.

A few details have been left open, particularly about object creation. The rules apply to routines that yield commands (procedures); their transposition to queries (functions and attributes, which return a result) needs to be made explicit.[61]

60. Semantic collaboration reuses the term "observer" from the Observer pattern, but with a specific meaning.

61. Also, *too many footnotes*.

## 12 Acknowledgments

This article is greatly indebted to the authors of the cited work about class-invariant-related issues. One should note, in considering its criticism of that work, that the first natural target for criticism is my own work in OOSC 1 and 2, which in its enthusiasm for the concept did not devote enough attention to the associated difficulties. More precisely it explained the reference leak problem in some detail (pages 407-410 in [23]) but not furtive access.

The development of AutoProof involved many discussions on the role of class invariants in verification; for the many insights gained I am grateful to the project members: Carlo Furia, Martin Nordio, Nadia Polikarpova, Julian Tschannen.

To advance the present work I gave several talks presenting intermediate states, prompting many important observations from the audiences. I cannot thank everyone but may note Sergey Velder at the PSSV 2016 symposium in Saint Petersburg, the organizers of that conference (Mikhail Itsykson and Nicolay Shilov); Philippe Quéinnec, Iulian Ober, Mamoun Filali, Jean-Paul Bodeveix, Jan-Georg Smaus, Sergei Soloviev and Peter Matthes at the University of Toulouse (in three "Vériclub" talks), Daniel de Carvalho, Alexander Chichigin and Alexander Naumchev at Innopolis University, Elisabetta di Nitto and Dino Mandrioli at Politecnico di Milano. Alexander Kogtenkov from Eiffel Software provided important comments on the first draft of this article. Special thanks are due to Iulian Ober from Toulouse for discovering a serious deficiency of the original version of the export-consistency rule in section 8.

For the furtive access problem, Daniel de Carvalho explored a notion of "aggregate invariant", which takes export status into consideration and led to useful discussions.

At one point during the course of her PhD thesis [40], Nadia Polikarpova mentioned that she was considering relying on export rights to address invariant-related problems; I am not sure why the idea was not pursued further.

## 13 References

[1] AutoProof page, with documentation, online tutorial, and references to publications, available at se.inf.ethz.ch/research/autoproof/.

[2] Ralph Back: *On Correct Refinement of Programs*, in *Journal of Computer and System Sciences*, vol. 23, no.1, pages 49-68, August 1981,

[3] Mike Barnett, David A. Naumann: *Friends Need a Bit More: Maintaining Invariants Over Shared State*, in MPC 2004, Mathematics of Program Construction, 7th International Conference, Stirling, Scotland, 12-14 July 2004, ed. Dexter Kozen, Lecture Notes in Computer Science 3125, Springer, 2004, pages 54-84.

[4] Mike Barnett, Robert DeLine, Manuel Fähndrich, K. Rustan M. Leino and Wolfram Schulte: *Verification of object-oriented programs with invariants*, Proceedings of FTfJP workshop at ECOOP 2004, in *Journal of Object Technology*, vol. 3, no. 6, 2004, pages 27-56.

[5] Mike Barnett (contact person): "Observer" verification challenge at SAVCBS workshop at ESEC/FSE conference 2007, www.eecs.ucf.edu/~leavens/SAVCBS/2007/challenge.shtml.

[6] Lilian Burdy, Yoonsik Cheon, David Cok, Michael Ernst, Joe Kiniry, Gary T. Leavens, K. Rustan M. Leino, and Erik Poll: *An overview of JML tools and applications*, in *International Journal on Software Tools for Technology Transfer*, vol. 7, no. 3, June 2005, pages 212-232.

[7] Ecma TC49-TG4 committee: *Eiffel: Analysis, Design and Programming Language*, Standard ECMA-367, 2nd edition, June 2006, available at www.ecma-international.org/publications/standards/Ecma-367.htm.

[8] Event-B page at www.event-b.org.

[9] Carlo Furia, Bertrand Meyer and Sergey Velder: *Loop invariants: Analysis, Classification and Examples*, in *ACM Computing Surveys*, vol. 46, no. 3, February 2014.

[10] Erich Gamma, Erich Gamma, Richard Helm, Ralph Johnson and John Vlissides: *Design Patterns: Elements of Reusable Object-Oriented Software*, Addison-Wesley, 1994.

[11] C.A.R. Hoare: *Procedures and Parameters: An Axiomatic Approach*, in *Symposium on Semantics of Algorithmic Languages*, ed. E. Engeler, Lecture Notes in Mathematics 188, Springer, 1971, pages 102-116.

[12] C.A.R. Hoare: *Proof of Correctness of Data Representation*, in *Acta Informatica*, vol. 1, no. 4, December 1972, pages 271–281.

[13] Alexander Kogtenkov, Bertrand Meyer and Sergey Velder: *Alias Calculus, Change Calculus and Frame Inference*, in *Science of Computer Programming*, 2015, pages 163-172, available at se.ethz.ch/~meyer/publications/aliasing/alias-scp.pdf.

[14] Gary Leavens and others: JML (Java Modeling Language) home page at www.jmlspecs.org.

[15] K. Rustan M. Leino and Peter Müller: *Object invariants in dynamic contexts*, in ECOOP 2004, Proc. 18th European Conference on Object-Oriented Programming, Oslo, 14-18 June 2004, ed. Martin Odersky, Lecture Notes in Computer Science 3086, Springer, pages 491-515.

[16] K. Rustan M. Leino and Peter Müller: *Modular verification of static class invariants*, in *FM 2005: Formal Methods, International Symposium of Formal Methods Europe*, Newcastle, July 18-22, 2005, eds. John Fitzgerald, Ian J. Hayes and Andrzej Tarlecki, Lecture Notes in Computer Science 3582, Springer, pages 26-42.

[17] K. Rustan M. Leino and Wolfram Schulte: *Using history invariants to verify observers*, in ESOP'07, Proc. 16th European Symposium on Programming, Springer, 2007, pages 80-94.

[18] K. Rustan M. Leino and Mike Barnett: Spec# home page at www.microsoft.com/en-us/research/project/spec/.

[19] Marx Brothers: *A Night at the Opera*, 1935. (The given citation starts at 3:54 at www.youtube.com/watch?v=G_Sy6oiJbEk.)

[20] Bertrand Meyer: *Eiffel: A Language for Software Engineering*, Technical Report TR-CS-85-19, Univ. of California, Santa Barbara, 1985, available at se.ethz.ch/~meyer/publications/eiffel/eiffel_report.pdf.

[21] Bertrand Meyer: *Object-Oriented Software Construction*, first edition, Prentice Hall, 1988.

[22] Bertrand Meyer: *Reusable Software: The Base Object-Oriented Component Libraries*, Prentice Hall, 1994.

[23] Bertrand Meyer: *Object-Oriented Software Construction*, second edition, Prentice Hall, 1997.

[24] Bertrand Meyer: *The Grand Challenge of Trusted Components*, in *ICSE '03: Proc. 25th Int. Conf. on Software Engineering*, Portland, Oregon, May 2003, IEEE Computer Society Press, 2003, pages 660-667, available at se.ethz.ch/~meyer/publications/ieee/trusted-icse.pdf.

[25] Bertrand Meyer: *The Dependent Delegate Dilemma*, in *Engineering Theories of Software Intensive Systems*, Proceedings of the NATO Advanced Study Institute on Engineering Theories of Software Intensive Systems, Marktoberdorf, Germany, 3 to 15 August 2004, eds. Manfred

Broy, J Gruenbauer, David Harel and C.A.R. Hoare, NATO Science Series II: Mathematics, Physics and Chemistry, vol. 195, Springer, June 2005.

[26] Bertrand Meyer: *Touch of Class: An Introduction to Programming Well Using Objects and Contracts*, Springer, 2009.

[27] Bertrand Meyer, Alexander Kogtenkov and Emmanuel Stapf: *Avoid a Void: The Eradication of Null Dereferencing*, in *Reflections on the Work of C.A.R. Hoare*, eds. C. B. Jones, A.W. Roscoe and K.R. Wood, Springer, 2010, pages 189-211, available at www.eiffel.org/doc-file/eiffel/void-safe-eiffel.pdf.

[28] Bertrand Meyer: *Verification as A Matter Of Course*, blog article with slides from a talk, 29 March 2010, available at bertrandmeyer.com/2010/03/29/verification-as-a-matter-of-course/.

[29] Bertrand Meyer, *Steps Towards a Theory and Calculus of Aliasing*, in *International Journal of Software and Informatics*, Chinese Academy of Sciences, 2011, pages 77-116, available at se.ethz.ch/~meyer/publications/aliasing/alias-revised.pdf.

[30] Bertrand Meyer et al.: SCOOP (Simple Concurrent Object-Oriented Programming) site, at www.eiffel.org/doc/solutions/Concurrent%20programming%20with%20SCOOP.

[31] Ronald Middelkoop, Cornelis Huizing, Ruurd Kuiper, and Erik J. Luit: *Invariants for non-hierarchical object structures*, in Electronic Notes in Theoretical Computer Science,195, 2008, pages 211–229.

[32] Caroll Morgan: *Programming from Specifications*, Prentice Hall, 1990-1998.

[33] Michal Moskal, Wolfram Schulte, Ernie Cohen and Stephan Tobies: *A Practical Verification Methodology for Concurrent Programs*, Microsoft Technical Report MSR-TR-2009-2019, 2009.

[34] Peter Müller: *Modular Specification and Verification of Object-Oriented Programs*, PhD thesis, Fernuniversität Hagen, 2001, Lecture Notes in Computer Science 2262, Springer, 2002.

[35] James Noble, David Clarke and John Potter: *Object Ownership for Dynamic Alias Protection*, in *TOOLS Pacific*, Melbourne, November 1999.

[36] Martin Nordio, Cristiano Calcagno, Peter Müller, Julian Tschannen and Bertrand Meyer: *Reasoning about Function Objects*, in TOOLS Europe 2010, Málaga (Spain), 28 June - 2 July 2010, ed. J. Vitek, Lecture Notes in Computer Science, Springer, 2010, available at se.ethz.ch/~meyer/publications/proofs/agents.pdf.

[37] David Parnas, *On the criteria to be used in decomposing systems into modules*, in *Communications of the ACM*, vol. 15 no. 12, December 1972, pages 1053-1058.

[38] Nadia Polikarpova, Carlo A. Furia, Yi Pei, Yi Wei and Bertrand Meyer: *What Good are Strong Specifications?*, in *Proceedings of ICSE 2013* (35th International Conference on Software Engineering), San Francisco, May 2013, IEEE Computer Press, pages 262-271, 2013, available at se.ethz.ch/~meyer/publications/methodology/strong_specifications_icse.pdf.

[39] Nadia Polikarpova, Julian Tschannen, Carlo A. Furia and Bertrand Meyer: *Flexible Invariants Through Semantic Collaboration*, in FM 2014 (proceedings of 19th International Symposium on Formal Methods), Singapore, May 2014, Lecture Notes in Computer Science 8442, eds. C. Jones, P. Pihlajasaari and J. Sun, Springer, 2014, pages 514-530, available at se.ethz.ch/~meyer/publications/proofs/flexible_invariants.pdf.

[40] Nadia Polikarpova: *Specified and Verified Reusable Components*, PhD thesis, ETH Zurich, available at se.ethz.ch/people/polikarpova/thesis.pdf.

[41] Mary Shaw, Ralph L. London and William A. Wulf: *An Introduction to the Construction and Verification of Alphard Programs*, in *IEEE Transactions on Software Engineering*, vol. 2, no 4, 1976, pages 53–265.

[42] Julian Tschannen, Carlo A. Furia, Martin Nordio and Bertrand Meyer: *Automatic Verification of Advanced Object-Oriented Features: The AutoProof Approach*, in *Tools for Practical Software Verification*; International Summer School, LASER 2011, eds. Bertrand Meyer and Martin Nordio, Lecture Notes in Computer Science 7682, Springer, December 2012. Other papers on AutoProof are listed in [1].

[43] Wikipedia, "class invariant" entry, last consulted July 2016.

[44] Akinori Yonezawa, Jean-Pierre Briot and Etsuya Shibyama: *Object-Oriented Concurrent Programming in ABCL/1*, in Proc. OOPSLA '86, Object-Oriented Programming Systems, Languages and Applications. ACM SIGPLAN Notices, vol. 21, no. 11, November 1986.